\documentclass[sigconf]{acmart}

\AtBeginDocument{%
  \providecommand\BibTeX{{%
    \normalfont B\kern-0.5em{\scshape i\kern-0.25em b}\kern-0.8em\TeX}}}

\setcopyright{acmcopyright}
\copyrightyear{2018}
\acmYear{2018}
\acmDOI{XXXXXXX.XXXXXXX}

\acmConference[Conference acronym 'XX]{Make sure to enter the correct
  conference title from your rights confirmation emai}{June 03--05,
  2018}{Woodstock, NY}
\acmPrice{15.00}
\acmISBN{978-1-4503-XXXX-X/18/06}

\usepackage{xspace}
\usepackage{soul}
\usepackage{fontawesome5}
\usepackage{graphicx}
\usepackage{subcaption}
\usepackage{tabularx}
\usepackage{multirow}
\usepackage{multicol}
\usepackage{enumitem}
\usepackage{longtable}
\usepackage{caption}
\newcommand{\zhenghighlight}[1]{\textcolor{orange}{TODO: }}

\newcommand{\crhighlight}[1]{{\textcolor{black}{#1}}}

\newcommand{\argu}{{\emph{Argumentative writing}}\xspace}

\newcommand{\visar}{{VISAR}\xspace}

\definecolor{figma_green}{HTML}{5A8F64}
\definecolor{figma_blue}{HTML}{6F91C3}
\definecolor{figma_orange}{HTML}{B76B56}
\definecolor{figma_yellow}{HTML}{C29735}

\copyrightyear{2023}
\acmYear{2023}
\setcopyright{acmlicensed}\acmConference[UIST '23]{The 36th Annual ACM Symposium on User Interface Software and Technology}{October 29-November 1, 2023}{San Francisco, CA, USA}
\acmBooktitle{The 36th Annual ACM Symposium on User Interface Software and Technology (UIST '23), October 29-November 1, 2023, San Francisco, CA, USA}
\acmPrice{15.00}
\acmDOI{10.1145/3586183.3606800}
\acmISBN{979-8-4007-0132-0/23/10}

\begin{document}

\title{VISAR: A Human-AI Argumentative Writing Assistant with Visual Programming and Rapid Draft Prototyping}


\author{Zheng Zhang}
\affiliation{%
  \institution{University of Notre Dame}
  \city{Notre Dame}
  \state{IN}
  \country{USA}}
\email{zzhang37@nd.edu}

\author{Jie Gao}
\authornote{Work done as a visiting student at the University of Notre Dame.}
\affiliation{%
  \institution{Singapore University of Technology and Design}
  \city{Singapore}
  \country{Singapore}}
\email{gaojie056@gmail.com}

\author{Ranjodh Singh Dhaliwal}
\affiliation{%
  \institution{University of Notre Dame}
  \city{Notre Dame}
  \state{IN}
  \country{USA}}
\email{rdhaliwa@nd.edu}

\author{Toby Jia-Jun Li}
\affiliation{%
  \institution{University of Notre Dame}
  \city{Notre Dame}
  \state{IN}
  \country{USA}}
\email{toby.j.li@nd.edu}


\begin{abstract}

In argumentative writing, writers must brainstorm hierarchical writing goals, ensure the persuasiveness of their arguments, and revise and organize their plans through drafting. Recent advances in large language models (LLMs) have made interactive text generation through a chat interface (e.g., ChatGPT) possible. However, this approach often neglects implicit writing context and user intent, lacks support for user control and autonomy, and provides limited assistance for sensemaking and revising writing plans. To address these challenges, we introduce VISAR, an AI-enabled writing assistant system designed to help writers brainstorm and revise hierarchical goals within their writing context, organize argument structures through synchronized text editing and visual programming, and enhance persuasiveness with argumentation spark recommendations. VISAR allows users to explore, experiment with, and validate their writing plans using automatic draft prototyping. A controlled lab study confirmed the usability and effectiveness of VISAR in facilitating the argumentative writing planning process.
 
\end{abstract}







\begin{teaserfigure}
  \includegraphics[width=\linewidth]{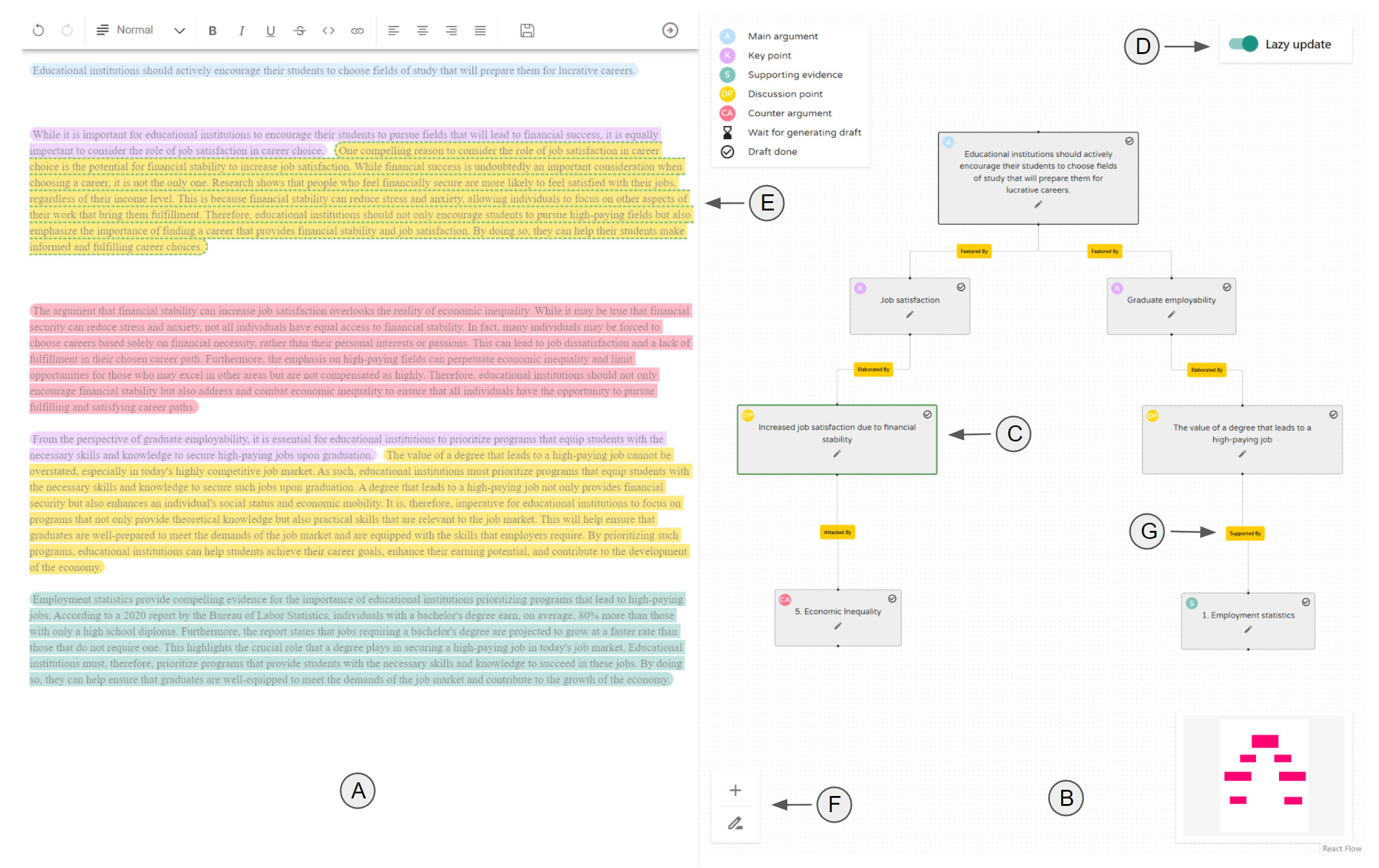}
  \caption{The main interface of \visar. The user can edit their argumentative writing outline via either the text editor (A) or the canvas of the visual outline (B); \visar synchronizes these two workspaces and facilitates users to correspond a writing item in editor (E) to the counterpart node in visual outline (C); User can switch between \textit{Lazy} or \textit{non-Lazy} mode to decide when to generate draft prototypes during the planning (D); User can choose between multiple relationships between items in outline (G); User can also add new item to the visual outline and generate a new draft from the outline (F). }
  \Description{The image shows the main interface of VISAR system}
    \label{fig:VISAR system}
\end{teaserfigure}

\maketitle

\section{Introduction}


Argumentative writing is a meta-genre of writing that is often used by individuals to persuasively convey their thoughts on disputable topics~\cite{blair2011groundwork, lunsford2016everything, williams_three_2012, gage_2006}. Unlike other forms and genres of writing, such as narrative and descriptive writing, argumentative writing requires writers to justify their claims with evidence and logical reasoning ~\cite{ferretti2018argumentative, blair2011groundwork, nelson1998theorizing}. Additionally, they must carefully identify and address potential counterarguments in order to construct a compelling argument~\cite{blair2011groundwork}. 

However, the argumentative writing process presents significant challenges, requiring considerable effort from writers. From a cognitive perspective, the challenge stems from the inherent conflict between two levels on which the writer must simultaneously operate. First, writers must create and maintain a hierarchical structure of arguments and sub-arguments to ensure global logical coherence and comprehensiveness of their argumentation~\cite{Kleemola2022TheCO}. This involves brainstorming constructs across various levels of abstraction, including high-level key aspects, specific discussion points and sub-arguments, evidence requirements, and potential counterarguments~\cite{blair2011groundwork}. Second, at the same time, writers must focus on the lower-level logical relationships between the discussion points to ensure their validity and coherence~\cite{ferretti2019argumentative}.  This writing process is both \textit{iterative} and \textit{non-linear} as specific content for discussion points and sub-arguments helps writers solidify their ideas when planning argument structure, while high-level structures support their efforts to maintain global logical coherence and comprehensiveness when working on specific texts~\cite{lovejoy2011great}.\looseness=-1


Although several interactive tools have recently been introduced to assist the argumentative writing process, especially with the advancement of large language models (LLMs) such as the GPT family~\cite{brown2020language}, these tools often focus on one of the two levels without effectively connecting them. For example, at the higher level of ideation, Sparks~\cite{gero2022sparks} facilitates ideation for scientific writers by using an LLM to suggest potential ideas related to given scientific concepts. Similarly, CoAuthor~\cite{lee2022coauthor} provides writers with context-based argumentative suggestions, inspiring them with the next piece of content to write. At the lower level of concrete text, some systems prioritize giving feedback on pre-existing drafts to help writers recognize weaknesses and adhere to argumentative structure in order to improve their writing~\cite{wambsganss2020adaptive, wambsganss2021arguetutor, almasri2019intelligent}. While these tools have proven useful for argumentative writers, they typically address specific writing constructs within a particular phase of the writing process (e.g., ideation, drafting, and revision) or a specific level of abstraction (e.g., global argumentative structures or local writing suggestions). Consequently, there remains an unmet challenge in developing effective methods to assist argumentative writers in generating and structuring their ideas through hierarchical writing planning and in seamlessly transitioning between global and local levels of abstraction across planning and drafting phases.\looseness=-1



To address these gaps, we introduce VISAR\footnote{VISAR stands for \textbf{V}isual \textbf{I}nteractive \textbf{S}ystem for \textbf{A}rgumentative writing with \textbf{R}apid draft prototyping}\footnote{Live demo website: \href{https://visar.app}{https://visar.app}}, a human-AI collaboration system that supports writers in the hierarchical and iterative planning process of argumentative writing.

\visar incorporates two novel approaches: \textit{visual programming} and \textit{rapid prototyping} for argumentative writing.  The interface of \visar features side-by-side text editors and an interactive graph of logical relationships among entities for writing planning. During the ideation process, \visar employs a state-of-the-art LLM to assist writers in interactively exploring potential discussion points using a ``chain of thought'' approach~\cite{wei2022chain} \crhighlight{(Figure \ref{fig:elaboration_flow})}. Entities and their relationships are visualized in a tree structure that users can freely manipulate and edit \crhighlight{(Figure \ref{fig:visual_manipulation})}. Entities are connected through relationships, and writers can add, edit, or remove entities and connect them using various relationship types in a visual programming interface. \visar also helps writers identify potential counterarguments, supporting evidence, and logical flaws for entities in the current argumentative plan \crhighlight{(Figure \ref{fig:argumentative_flow})}, synchronizing both the visualization and the text editor. This method allows writers to monitor and evaluate the ideation process across different abstraction levels and facilitates the gradual refinement of the language model's thoughts in a step-by-step manner, enabling it to evolve its context understanding and provide context-consistent suggestions that respond to writers' needs.

\visar's rapid prototyping approach generates concrete text of what the final write-up of the argumentation \textit{could be} based on the current argumentation structure using state-of-the-art LLMs. It is important to note that \visar's objective is \textit{not} to produce the final write-up, as LLMs still have limitations in, e.g., achieving the desired levels of factual correctness and specificity. Instead, \visar aims to generate \textit{mid-fidelity} (hi-fidelity in form and logical flow; low-fidelity in factual details) prototypes of the write-up to facilitate fast and iterative exploration of ideas. Writers first ``diverge'' by selecting different key points, argumentation structures, and supporting points, and then ``converge'' by examining the generated write-up prototypes to understand the strengths, weaknesses, and trade-offs of various approaches, reaping the benefits of \textit{parallel prototyping} in user experience design~\cite{lunzer_2008_subjective}.



The design of \visar embodies our vision of effective human-AI collaboration on complex and ambiguous tasks with multiple intertwined layers of abstraction, particularly in an era when LLMs like GPT-4 are more powerful than ever. Recently, interfaces that wrap LLMs into multi-turn chatbots, such as ChatGPT, have gained considerable attention. Users find them capable of generating \textit{seemingly} reasonable argumentative essays from simple prompts (e.g., ``Can you write an argumentative essay about...'') or auto-completion. Furthermore, the ability to maintain context in multi-turn conversations allows them to support \textit{highly flexible} follow-up requests on both language style (e.g., ``Can you revise this section in the style of academic writing?'') and content (e.g., ``Can you come up with a specific example for this?'') levels. However, we argue that the chat interface is \textit{not} the ultimate solution for tasks like this. As evidenced by the influential Direct Manipulation vs. Agent debate~\cite{shneiderman_1997_direct} over two decades ago, a direct manipulation interface offers several key advantages over agents, including better transparency into the system's state, easier error handling, finer granularity of user control, useful constraints to guide user actions, and more visible system affordances. The design of \visar incorporates these guidelines by introducing a visual outline of the argumentation structure and enabling object-level direct manipulation (selecting, dragging, actions) on both visual and text representations. In designing \visar, our goal is to retain the flexibility provided by LLMs while offering users more control, constraints, and transparency to guide users through the task.


The results of a controlled lab evaluation showed that users were able to successfully use \visar to assist with their argumentative writing planning process. Participants found the various intelligent features and interaction strategies in \visar useful and expressed interest in incorporating \visar into their own argumentative writing planning process. 

In summary, this paper presents the following contributions:

\begin{itemize}
    \item a new approach that uses visual programming and rapid prototyping strategies to achieve effective collaboration between human writers and LLMs with adequate user control and autonomy in the ideation and planning stages of argumentative writing.
    
    \item \visar, a writing assistance tool that implements this approach to support writers in argumentative writing tasks.
    
    \item a within-subjects user study with 12 participants that validated the usability and effectiveness of \visar.
\end{itemize}

\section{Related Work}

\subsection{Writing Support Tools}

The development of tools to support users in the writing process has been a long-standing area of interest. On the commercial side, tools for grammar and spelling check, like the built-in tool of Microsoft Word and third-party tools (e.g., Grammarly\footnote{\url{https://www.grammarly.com/}}) predominantly address language correctness without providing guidance on content.  Beyond grammar and spelling, tools like Ref-N-Write\footnote{\url{https://www.ref-n-write.com/}} expand on this by focusing on language style, helping users paraphrase their writing to achieve more concise, professional, or academic expressions. 
Ref-N-Write also offers practical suggestions for developing research paper outlines and drafts, as well as providing diverse inspirations for academic writing in various fields \cite{ref-n-write_2023}.

The incorporation of natural language generation (NLG) technology into writing support tools has become increasingly prevalent in recent years. Commercial platforms, such as the online AI-powered Smodin\footnote{\url{https://smodin.io/}}, assist users with rewriting and employ text generation technology to produce essay outlines for different genres, including argumentative and descriptive writing.


In the academic domain, extensive research has been conducted on general writing support. \crhighlight{For creative writing, Kreminski et al. \cite{kreminski2022unmet} investigated the need for writers, including maintaining narrative consistency, developing plot structure, crafting reader experiences, and refining expressive intent.} To support human-AI \crhighlight{collaboration} for \textit{creative writing} (as opposed to argumentative writing), Coenen and Ippolito et al. \cite{ippolito2022creative, coenen2021wordcraft} introduced Wordcraft, a story-writing tool that rewrites, elaborates, or provides personalized prompts for subsequent text. CoAuthor \cite{lee2022coauthor} is an interactive writing tool that leveraged the GPT-3 model to offer writers ideas for their next writing steps in both creative and argumentative writing, while Chakrabarty et al. developed CoPoet \cite{chakrabarty2022help}, a collaborative poetry tool system based on LLMs that generates poetry in response to user-provided text and customized prompts. Yang et al. \cite{yang2022ai} explored interaction strategies used by human writers to condense, revise, summarize, and regenerate AI-generated text in fictional writing. The work of Singh et al.~\cite{singh2022hide} explored a multi-modal technique that visualizes story plots to support ideation in creative writing. An empirical study~\cite{bhat2023interacting} by Bhat et al. suggested that when writers encounter next-phase suggestions from an LLM, they take aid from them in multiple nuanced ways even if they do not directly accept the suggestions.

For tools that support specific parts of the argumentative writing process, Dang et al. \cite{dang2022beyond} proposed a text editor that generates continuous text summaries alongside the main text, which was shown useful in helping users revise the content and scope of their drafted paragraphs. For final reflection and quality improvement, Wambsganss et al. \cite{wambsganss2020adaptive} introduced AL, an \argu support system that provides feedback on argument quality by identifying argument components (e.g., claims, premises, non-argumentative elements) and recognizing argument relations. Xia et al. \cite{xia2022persua} presented Persua, which employs machine learning models to detect components of argumentative writing and their relationships, and provides example-based guidance on persuasive techniques to improve persuasiveness of arguments. Wong et al.~\cite{wong2015tabletop} developed a tabletop system to promote argumentation in computer science students. On the machine learning research side of augmenting the argumentation capabilities of LLMs, Yang et al. \cite{yang2022re3} introduced $\mathbf{Re^3}$, a Recursive Reprompting and Revision framework for automatically generating longer stories by iteratively creating new prompts using intermediate information.

Our work with \visar complements the existing literature by focusing on interaction strategies and collaborative workflows for human-AI collaboration in two crucial early stages of argumentative writing: prewriting and planning. \visar proposes a new workflow that employs visual programming and rapid prototyping methodologies to ensure user flexibility, autonomy, and control within the established process of argumentative writing.\looseness=-1 

\subsection{Human-AI Collaboration in Creative Work}

\crhighlight{ Cognitive theory \cite{kozbelt2010theories} models creative work as a four-stage process: \textit{preparation, incubation, illumination, and evaluation}. In addition, creativity is said to involve two thinking modes: \textit{divergent} and \textit{convergent} thinking~\cite{muller2011leaving, kahneman2011thinking}. Divergent thinking is most commonly associated with the process of generating original ideas and breaking away from conventional ways of thinking. Once a broad range of ideas has been generated through divergent thinking, convergent thinking can be used to evaluate, refine, and implement these ideas.} 

\crhighlight{Recently, there are some works exploring to use AI models to support people's workflow of creative work~\cite{ippolito2022creative, kreminski2022unmet, lu_bridging_2022}. Examples include collaborative story writing tools such as TaleBrush \cite{chung2022talebrush}, which allows users to create stories with AI assistance through sketching interactions, indicating different story stages such as positive or negative events. TaleBrush primarily aids the planning stage of writing, enabling writers to generate diverse story lines and refine them iteratively. In a different context, Storybuddy \cite{zhang2022storybuddy} uses question-answer generation (QAG) to create question-and-answer pairs based on the context of a story, facilitating interactive storytelling between parents and children. Similarly, Lee et al. \cite{lee2022interactive} suggested using AI to collaboratively rewrite stories for children and parents, while Akoury et al. \cite{akoury2020storium} presented STORIUM, an evaluation platform that allows users to request language model suggestions during story writing. In addition, CodeToon\crhighlight{~\cite{suh2022codetoon}} is an AI-powered story ideation and auto-comic generation tool that maps from code}.

\crhighlight{Applying AI to creative work presents unique challenges. Creative work requires more than just the spark of an idea---it must exhibit originality and effectiveness \cite{runco2012standard} of the content creator. While AI models, especially large language models, are good at producing diverse and in-context content on the fly, ensuring the originality and authenticity of their outputs is challenging. For example, generative AI models are known to have tendency to produce outputs that deviate from reality, often referred to as ``hallucination''~\cite{kaddour2023challenges}. Users must take responsibility for verifying the credibility of the sources of the generated content. Additionally, the authorship of AI-generated creative work might spark controversy as well~\cite{ballardini2019ai}}. \looseness=-1

\crhighlight{Our work focuses on employing AI to streamline the initial prewriting and planning stages of argumentative writing. The goal is to inspire writers and generate early drafts that serve mainly to enhance understanding and make sense of complex writing topics, rather than incorporating AI-generated content in the final product \cite{gao2023collabcoder}, a practice that could raise ethical concerns.}


\subsection{Prompting and Prompting Chain in LLMs}

\visar interacts with its underlying large language model (LLM) through \textit{prompting} \cite{liu2023pre}. Specifically, a new diagram, ``pre-train, prompt, and predict'' \cite{liu2023pre}, has recently been proposed. To achieve better performance, diverse downstream tasks are reformulated to resemble the type of task the model was originally trained on, rather than adapting the model's objective function to the downstream task. Hence, LLMs are often considered ``general purpose'' pre-trained models.\looseness=-1


To guide an LLM in generating results, various prompting strategies can be employed \cite{reynolds2021prompt}. Zero-shot prompting involves a question-answer mechanism, while one-shot prompting provides one example for the AI to base its output on. Few-shot prompting involves providing the AI model with multiple examples, and role prompting requires the AI to complete a task based on a given role rather than examples or templates. Notably, Wei et al. \cite{wei2022chain} introduced the chain of thought (CoT) to elicit reasoning from the model. This technique enables models to break down complex, multi-step problems into intermediate steps by offering a sequence of prompts that guide them towards a final objective. Implementing CoT significantly enhances the performance of previous language models in reasoning tasks previously considered ``off-limits'' for language models, even achieving human-parity performance in some tasks \cite{suzgun2022challenging}.


Argumentative writing, as described in Section \ref{sec:background}, can be decomposed into several steps, such as proposing issues, claims, and evidence. Importantly, writing is an iterative process that requires writers to continuously refine their thoughts. This process bears a resemblance to the concept of CoT. In fact, Yang et al. \cite{yang2022re3} demonstrated that the $Re^3$ framework can generate longer stories with a CoT approach that implements recursive reprompting and revision. In this paper, \visar uses a human-in-the-loop prompt chain approach for argumentative writing. By combining the strengths of CoT and human expertise, our approach aims to enhance the effectiveness of human-AI collaboration in argumentative writing.

\section{Background of Argumentative Writing}
\label{sec:background}
Argumentative writing is a type of writing in which the writer presents and supports a position on a controversial issue using evidence and reasoning, with the goal of persuading the reader to accept the viewpoint of the writer or take a specific action \cite{gage_2006, blair2011groundwork, lunsford2016everything, williams_three_2012}. This writing style relies on arguments as foundational components, where each argument posits a statement on an issue, substantiated with evidence and reasoning. Multiple arguments can be used to build a larger argument or present a comprehensive perspective on a subject.


Toulmin's argument model \cite{hitchcock2006arguing} provides a framework for formulating logical arguments. It consists of six components:  \textit{claim} (the main point to be supported), \textit{data} (evidence validating the claim), \textit{warrant} (the underlying assumption or rationale linking the data to the claim), \textit{backing} (supplementary evidence or reasoning supporting the warrant and strengthening the argument), \textit{qualifier} (a statement limiting the claim's scope or force), and \textit{rebuttal} (counterarguments or objections to the claim). This model underscores the necessity of evidence and reasoning to reinforce claims while anticipating and addressing potential objections or counterarguments. \crhighlight{Grounded in theoretical argumentation models, argument mining~\cite{wachsmuth2017computational, lawrence2020argument} seeks to autonomously detect and extract the structure of inferential logic and reasoning articulated as arguments in natural language. This technique aids in evaluating individuals' stances and the underlying rationales they employ, helping writers identify the structural and logical weakness in their arguments.} Informed by Toulmin's model \crhighlight{and argument mining works}, \visar aims to prompt writers about how to strengthen arguments by alerting them to potential counterarguments (rebuttals), supplying supporting evidence (data, backing), and scrutinizing logical fallacies primarily related to claim, warrant, and qualifier. \crhighlight{Compared with traditional argument mining approaches, using LLM to inform argument structure has advantages such as adaptability across various domains, obviating the requirement for dataset collection, and enhanced accuracy in contextual comprehension. }

Drawing on Toulmin's model, \visar aims to assist writers in strengthening their arguments by alerting them to potential counterarguments (rebuttals), providing supporting evidence (data, backing), and examining logical fallacies, primarily concerning the claim, warrant, and qualifier.

\begin{figure}[!t]
  \centering
  \includegraphics[width=\columnwidth]{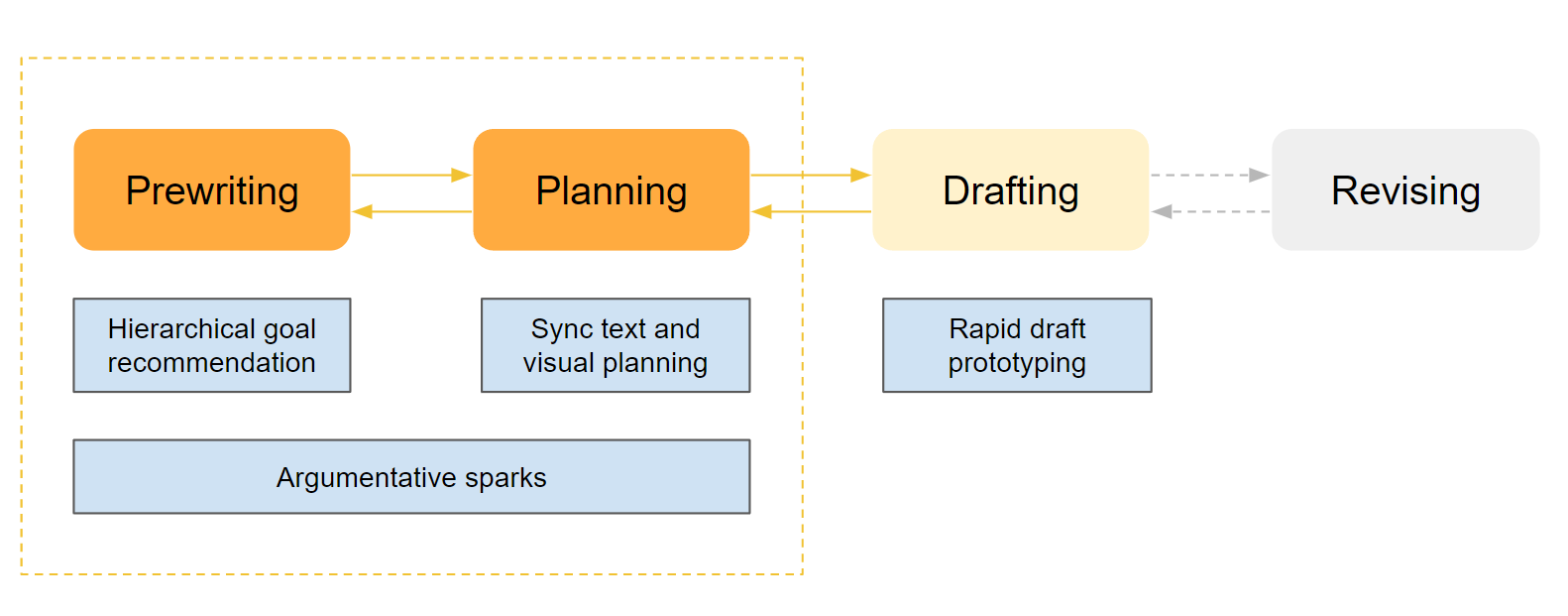}
  \caption{Four cognitive writing stages~\cite{flower1981cognitive, grabe2014theory} and how \visar's features support them. \visar primarily focuses on \textit{prewriting} and \textit{planning} stages in argumentative writing. \visar also assists with rapid draft prototyping as a way to facilitate more effective iterations in the previous two stages.}
 
  \label{fig:writing_process_feature_support}
  \vspace{-3mm}
\end{figure}


Argumentative writing can be categorized into two types based on the direction of reasoning \cite{pedemonte2007can, schwarz2017dialogue, nelson1998theorizing, booth2003craft}: \textit{deductive} and \textit{inductive}. \textit{Inductive argumentation}, a \textit{bottom-up} approach, starts with specific observations or instances and progresses toward broader generalizations or theories. Inductive reasoning examples include scientific, statistical, and analogical argumentation. Conversely, \textit{deductive argumentation}, a \textit{top-down} process, begins with general principles or premises and proceeds to specific conclusions through logical reasoning. This approach involves starting with an established theory or hypothesis and using it to derive particular implications, predictions, or conclusions.

\textbf{\visar focuses on facilitating deductive, top-down argumentative writing.} Compared to inductive arguments, deductive arguments are more common in written communication~\cite{johnson1999deductive}. In situations such as decision making, problem solving, or expressing opinions, the top-down strategy enables individuals to communicate their ideas coherently and persuasively  \cite{johnson1999deductive}.\looseness=-1

Theories of the writing process \cite{flower1981cognitive, grabe2014theory} identify four primary cognitive stages of the writing process: \textit{prewriting, planning, drafting, and revising} (Figure~\ref{fig:writing_process_feature_support}). \textbf{\visar primarily concentrates on assisting writers during the prewriting and planning stages.} However, to support the iterative and non-linear nature of the writing process, we also enable users to rapidly generate draft prototypes, allowing them to review and revise their current outlines within a tangible writing context.
In the prewriting stage of deductive writing, writers gather information and brainstorm hierarchical writing objectives, including high-level key aspects, discussion points, and supporting evidence. This stage can generate a high cognitive load, as writers must retrieve information and sequentially elaborate their thought process. In the planning stage, developing a hierarchical and logical structure for the argumentation outline can be challenging in deductive writing. Writers must ensure that their premises logically lead to the intended conclusion with necessary evidence and settlement of counterarguments,  while maintaining coherence and consistency throughout the planning.

Here, we define the key terms we use throughout this paper:

\begin{itemize}
    \item \textbf{Argument}: A statement or group of statements called premises intended to determine the degree of truth or acceptability of another statement called a conclusion \cite{johnson2012manifest}.
    \item \textbf{Argumentation}: A process of constructing, presenting, and defending a position or claim through the use of logical reasoning, evidence, and persuasive techniques \cite{blair2011groundwork}.
    \item \textbf{Writing goal}: An intended outcome that a writer seeks to achieve while composing an argumentative piece in order to persuade audiences to accept the central claim \cite{flower1981cognitive}.
    \item \textbf{Key aspect}: A primary theme or central concept that guides the development and organization of argumentation to support the central claim from a particular perspective \cite{flower1981cognitive}.
    \item \textbf{Discussion point}: A specific point or idea that supports the central claim related to a particular theme \cite{flower1981cognitive}.
    \item \textbf{Counter argument}: An opposing viewpoint or objection that challenges the central claim, or position presented by the writer \cite{hitchcock2006arguing}.
    \item \textbf{Evidence}: Information, facts, data, or examples that support the writer's central claim or argument \cite{hitchcock2006arguing}.
    \item\textbf{Logical fallacy}: The use of flawed or unsound reasoning in the development of an argument, which may appear to be a well-reasoned argument if unnoticed \cite{blair2011groundwork}.
    \item \textbf{Textual outline}: A structured text-based framework that organizes the writing components such as main ideas, arguments, counterarguments, and supporting evidence. An example is nested bullet points \cite{lindgren2006writing}.
    \item \textbf{Visual outline}: A graphical representation of the structure and organization of argumentation that highlight the logical connections among components. Examples include mind maps, flowcharts, and hierarchical diagrams \cite{lindgren2006writing}.
\end{itemize}




\section{VISAR system}

\begin{figure*}[htp]
  \centering
  \includegraphics[width=\linewidth]{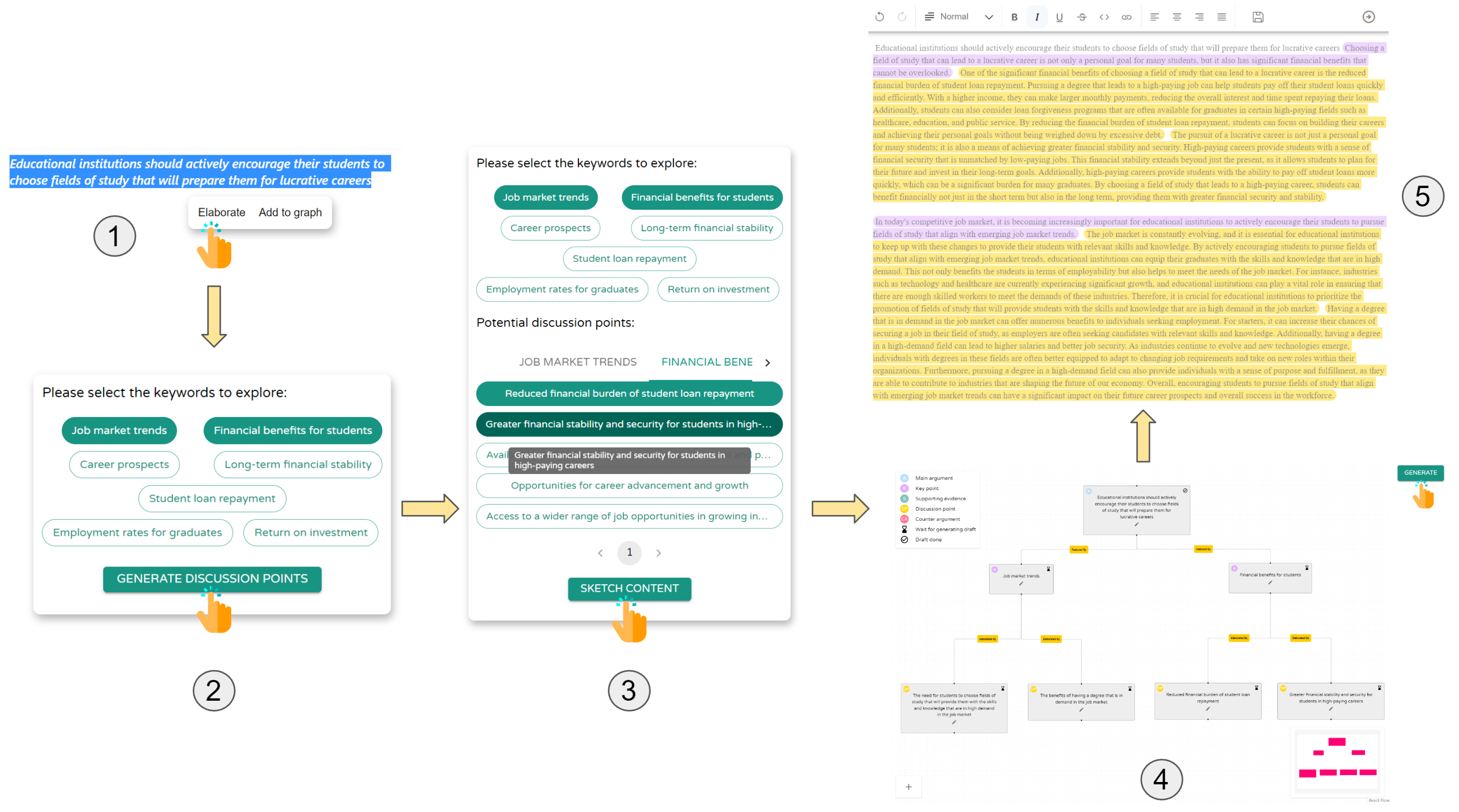}
  \caption{The interaction flow of hierarchical goal recommendation: (1) Users can select a range of text and click \textit{Elaborate} to invoke hierarchical goal recommendation; (2) \visar proposes a set of possible key aspects, users can select key aspects that they want to further explore; (3) \visar proposes specific discussion points for each selected key aspect; (4) Users can review and modify the the logical structure of the writing goals. By click \textit{Generate}, users can use \visar to implement the draft for the current writing outline; (5) \visar generates a prototype of the draft in the text editor.}
 
  \label{fig:elaboration_flow}
  \vspace{-3mm}
\end{figure*}

\subsection{Design Goals}


\crhighlight{ Our goal is to develop a human-AI collaborative system that facilitates writers in the hierarchical and iterative planning process of argumentative writing. To inform our system design, we conducted a semi-systematic literature review using method proposed by Synder ~\cite{snyder2019literature} to gain insights into (1) the cognitive processes of argumentative writing, (2) the challenges faced by argumentative writers, and (3) the capabilities and limitations of existing writing assistance tools. Specifically, we first identified seminal and extensively referenced works in the fields of cognitive writing theory, argumentation theory, writing pedagogy theory and writing assistance tools. This step helped us gain a fundamental understanding of the aforementioned questions. Afterwards, we expanded our inquiry to more recent works in those fields  via keyword searches and by tracing the citation networks from the key works recognized in the preceding stage. This allowed us to assimilate recent advancements in these domains as well as understand the persisting challenges. Through the process of this literature review, we identified four design goals that our system ought to achieve:} \looseness=-1


\paragraph{\textbf{DG1: Accommodating the non-linear and hierarchical nature of writing planning process}}

Rhetorical theories often conceptualize writing as a goal-directed process, in which writers must establish a hierarchical network of goals to guide their efforts \cite{flower1981cognitive, nelson1998theorizing}. In argumentative writing, these objectives involve introducing the topic and main argument, elaborating on key points and sub-arguments supporting the central argument, providing evidence for the same, and addressing potential counterarguments. However, users typically determine local goals only after engaging in a hierarchical planning process that provides a concrete context~\cite{lovejoy2011great}. Furthermore, instead of following a linear process, writers frequently revise goals or adjust their relationships as they gain deeper understanding during the planning phase \cite{gollins2016framework}. Due to the hierarchical structure of planning, these revisions may cascade and affect subordinate subgoals, making it difficult to maintain coherence \cite{ferretti2019argumentative}. To address these characteristics, the system should help writers construct a hierarchical sequence of writing goals while concurrently aiding them in maintaining overall coherence during goal revision.


\paragraph{\textbf{DG2: Facilitating the ideation of writing plans}}

Writing sometimes begins as a serendipitous experience. Writers may start planning without a clear endpoint or a comprehensive understanding of required components \cite{flower1981cognitive, ferretti2019argumentative}. Additionally, the writing planning process requires writers to quickly retrieve relevant information from long-term memory in accordance with the current local goal \cite{flower1981cognitive}. However, this information may be underdeveloped or disorganized, which can lead to omissions \cite{wentzel2017guide}. In response, our proposed system should provide writers with diverse options at each ideation stage, inspiring possible subsequent writing goals based on the existing context.


\paragraph{\textbf{DG3: Scaffolding to enhance persuasiveness of users' argumentation}}

Argumentative writing requires writers to justify their arguments' credibility through persuasive strategies (e.g., \textit{logos}, \textit{ethos}, \textit{pathos}) and addressing counterarguments and logical fallacies \cite{blair2011groundwork, ruiz2022qualitative}. However, maintaining persuasiveness can be challenging. Writers may be unfamiliar with persuasive strategies and common types of logical fallacy or struggle to select and implement suitable strategies for a specific context. Therefore, our system should prompt writers with relevant persuasive strategies, highlight potential logical fallacies, and present counterarguments tailored to the specific local context in which the writers operate.

\paragraph{\textbf{DG4: Facilitating users' reflection and assessment of plans}}

The planning and drafting stages are often intertwined in the writing process. As writers begin to draft, they may realize that some aspects of their outline are incomplete or lack sufficient detail, possibly due to missing evidence, weak arguments, or unclear transitions between sections \cite{flower1981cognitive, grabe2014theory}. Drafting allows writers to identify gaps, return to the planning stage, and revise their outline accordingly \cite{flower1981cognitive, ferretti2019argumentative}. It also enables writers to reassess ideas and restructure their plan. However, transforming ideas from planning into a coherent, well-structured piece of writing can be time-consuming and laborious. To alleviate these efforts, our system should support users in reflecting on and evaluating their plans through automated rapid draft prototyping.

\begin{figure*}[htp]
  \centering
  \includegraphics[width=\linewidth]{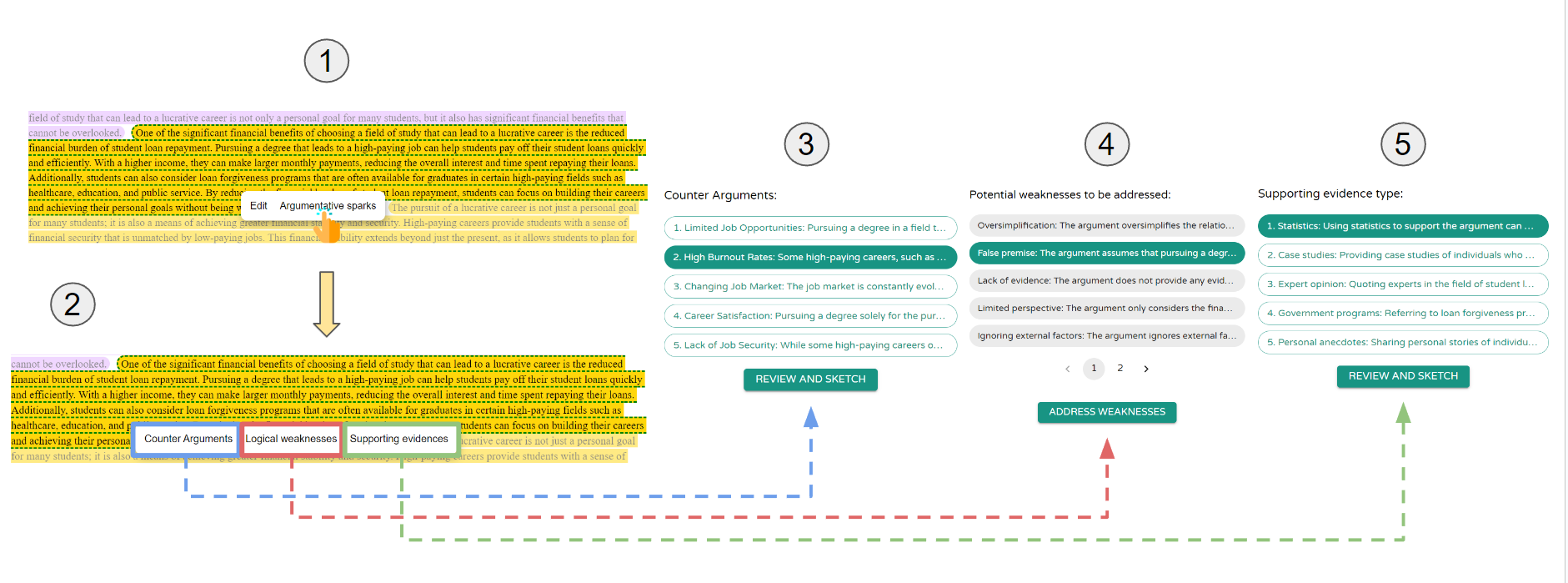}
  \caption{The interaction flow of argumentative sparks: (1) users can invoke argumentative sparks by clicking on a text block; (2) users can select the specific spark type; (3) \visar proposes counterarguments for the selected discussion point; (4) \visar identifies logical fallacies in the selected content; (5) \visar recommends supporting evidences for the selected content.}
 
  \label{fig:argumentative_flow}
  \vspace{-3mm}
\end{figure*}

\subsection{Example scenario}

In this section, we provide an illustrative example of how \visar operates in a practical context. Imagine that Alice needs to write an argumentative essay supporting the statement ``\textit{Universities should require every student to take a variety of courses outside the student's field of study}''. Unsure of the best structure and needing guidance to create a persuasive argument, Alice decides to use \visar to plan and draft her essay.

Alice begins by entering the statement as her initial sentence in \visar (users may also choose to write more themselves before seeking AI assistance). After highlighting the text and clicking the \textit{Elaborate} button (Figure \ref{fig:elaboration_flow}-1), \visar provides Alice with step-by-step writing goal recommendations. First, it suggests key aspects of the argument that Alice may consider, such as \textit{breadth of knowledge, well-rounded education, career preparation, and development of critical thinking skills}. Alice selects \textit{well-rounded education} and \textit{career preparation} for further exploration.

By clicking the \textit{Generate discussion points} button (Figure \ref{fig:elaboration_flow}-2), \visar generates potential points related to the two selected aspects. For \textit{well-rounded education}, it prompts Alice to consider the \textit{benefits of exposure to diverse thinking styles} and the \textit{opportunities for lifelong learning and personal growth}. For \textit{career preparation}, \visar recommends addressing \textit{preparation for interdisciplinary jobs} and \textit{increased adaptability in a fast-changing job market}. Alice selects the points of lifelong learning and adaptability to the changing market and clicks the \textit{Sketch Content} button (Figure \ref{fig:elaboration_flow}-3) to review the current logical structure of her writing goals.

Alice then clicks the \textit{Generate} button (Figure \ref{fig:elaboration_flow}-4) to request prototype drafts for each goal in her outline, which she can directly edit. She also asks \visar to generate alternative drafts for the point on \textit{lifelong learning} and \textit{personal growth} and refines the draft through conversation, providing further instructions.

To identify counterarguments, logical fallacies, and necessary supporting evidence in her drafts, Alice uses \visar's Argumentative Sparks feature. She selects the draft block and clicks the \textit{Argumentative sparks} button (Figure \ref{fig:argumentative_flow}-1), which provides her with likely counterarguments, potential logical fallacies, and suggestions for supporting evidence. \visar also enables Alice to view draft implementations of counterarguments and supporting evidence, along with revised drafts of original discussion points that address logical weaknesses.

\begin{figure*}[t!]
  \centering
  \includegraphics[width=\linewidth]{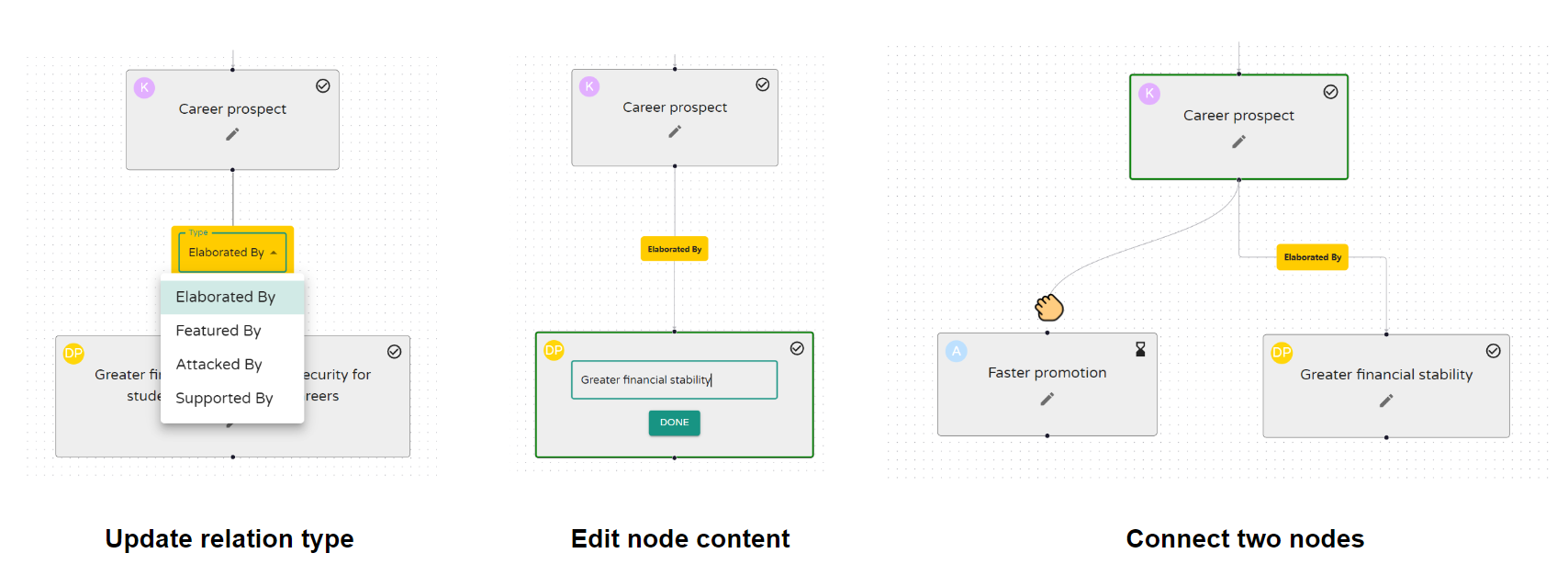}
  \caption{Illustration of three types of manipulations on the visual outline}
 
  \label{fig:visual_manipulation}
  \vspace{-3mm}
\end{figure*}

To generate a visual outline, Alice takes advantage of \visar's visual programming feature. A visual outline of the current argumentation outline is automatically generated, synchronized, and displayed next to the text editor.  Using the intuitive drag-and-drop interface (Figure \ref{fig:VISAR system}-B), she can add, edit, and rearrange goals within her outline (Figure \ref{fig:visual_manipulation}). For instance, Alice may want to include a new key aspect, \textit{development of critical thinking skills}, by creating a new node and connecting it as a child of the main argument node. As she modifies the visual outline, Alice can immediately view the updated draft in a pop-up window and update dependent nodes and goals in-situ. By activating the \textit{Lazy update} mode (Figure \ref{fig:VISAR system}-D), Alice can focus on editing the visual outline and update the entire draft simultaneously by clicking the \textit{Generate} text (\faPencil*) button (Figure \ref{fig:VISAR system}-F). Additionally, Alice can edit an argument or goal in the text editor and incorporate it into the visual outline by clicking the \textit{Add to graph} button (Figure \ref{fig:elaboration_flow}-1).

\subsection{Key Features}

This section introduces the key features of \visar, as illustrated in the example scenario. These features include hierarchical writing goal recommendation, synchronized text and visual planning, argumentative sparks,and automatic draft prototyping.\looseness=-1

\subsubsection{Hierarchical writing goal recommendation}
As noted in DG1 and DG2, writers often face challenges when generating a comprehensive set of potential goals at each planning level for their writing. \visar addresses this need by providing users with a step-by-step goal recommendation process, allowing them to direct goal generation at each level. The design of \visar facilitates the integration of user-defined and AI-generated goals, promoting user autonomy.

\paragraph{Inspiring writer with step-by-step writing goal recommendation}

Theories of the writing process \cite{flower1981cognitive, grabe2014theory} suggest that during the planning phase of writing an article, writers typically begin by brainstorming a wide range of relevant topics. They then identify specific points to support the core argument related to the chosen topics. To accommodate this process, \visar prompts writers in a ``chain of thought'' approach, allowing them to explore various potential topics connected to their core argument. \visar then provides recommendations for specific discussion points that can be further developed based on the selected topics.

To perform this step-by-step planning with an LLM, writers can select the text containing the main argument. By clicking the \textit{Elaborate} button in the displayed menu (Figure \ref{fig:elaboration_flow}-1), writers are presented with a set of suggested potential aspects or topics that can help substantiate the selected argument. Writers can review and select topics for further exploration. When writers click the \textit{Generate discussion point} button (Figure \ref{fig:elaboration_flow}-2), the system prepares specific discussion points organized by the selected topics.  Writers can navigate these by clicking the corresponding tabs and select discussion points to include in their outline (Figure \ref{fig:elaboration_flow}-3). After writers selected the points to discuss, \visar will show the current visual outline (see details in Section \ref{sec:visual_planning}) for writers to review the outlining structure. Writers can also click the \textit{Generate} button (Figure \ref{fig:elaboration_flow}-4) to produce a prototype write-up that implements this outline (see details in Section \ref{sec:draft_prototyping}).


\paragraph{Integration of user-defined goals with AI-generated goals}

\visar allows writers to interweave their own defined goals with AI-generated goals during the ideation process. Writers can enter a goal description in the text editor, select the range, and click the \textit{Add to graph} button to insert a new node in the visual outline that corresponds to the user-defined goal (see details in Section \ref{sec:visual_planning}). Writers can also create the node directly by clicking \textit{Add} (\faPlus) button (Figure \ref{fig:VISAR system}-F) and editing the node description. After adding the node, writers can position it appropriately in the visual outline by dragging (Figure \ref{fig:visual_manipulation}). Writers can collaborate with the LLM to further recursively elaborate their defined goals or any AI-generated discussion point by clicking the \textit{Elaborate} button on the corresponding text description.

\subsubsection{Synchronized text and visual planning}
\label{sec:visual_planning}

As highlighted in DS4, \visar supports both text-based and visual-based writing planning, catering to diverse planning strategies. This design combines the benefits of ``WYSIWYG (What You See Is What You Get)'' directness with text editing and the flexibility and clarity of organizing argumentation structure through visual programming.

\paragraph{Enabling interactive visual planning of writing}

\visar allows users to plan their writing by constructing a hierarchical tree-based outline on an interactive canvas that represents their writing outline. This visual representation is similar to existing visual programming and online whiteboard tools such as Miro\footnote{\url{https://miro.com/}}.  The nodes represent individual components or discussion points in the outline. The directed edges between nodes represent their relationships. Writers can easily create, remove, or edit nodes and edges through direct manipulation. In specific, \visar supports five types of nodes: \textit{main argument (default), key aspect, discussion point, counterargument and supporting evidence} and four types of edges: \textit{featured by} (child is a high-level topic/aspect regarding the parent), \textit{elaborated by} (child is a concrete discussion point regarding the parent), \textit{attacked by} (child is a counterargument that attacks the parent) and\textit{ supported by} (child is an evidence/argument supporting the parent). The node type will change accordingly as writers modify the edge type between it and its parent (Figure \ref{fig:visual_manipulation}). Writers can edit the contents of the nodes by clicking the \textit{Edit} (\faPen) button (Figure \ref{fig:VISAR system}-C).\looseness=-1 

\paragraph{Integrating text- and visual-based planning}

\visar enables writers to plan their writing through the correspondence between text-based and visual-based planning. Writers can initiate the writing process with text-based approaches such as listing (with bullet points) or freewriting, and integrate their arguments into a visual graph by selecting text and clicking the \textit{Add to graph} button (Figure \ref{fig:elaboration_flow}-1). To facilitate the correspondence of components between two representations, when an entity is selected in one representation, \visar highlights its counterpart in the other presentation. \visar also renders text blocks in the same color as their corresponding nodes to reinforce the correspondence.

\begin{figure}[!t]
  \centering
  \includegraphics[width=\columnwidth]{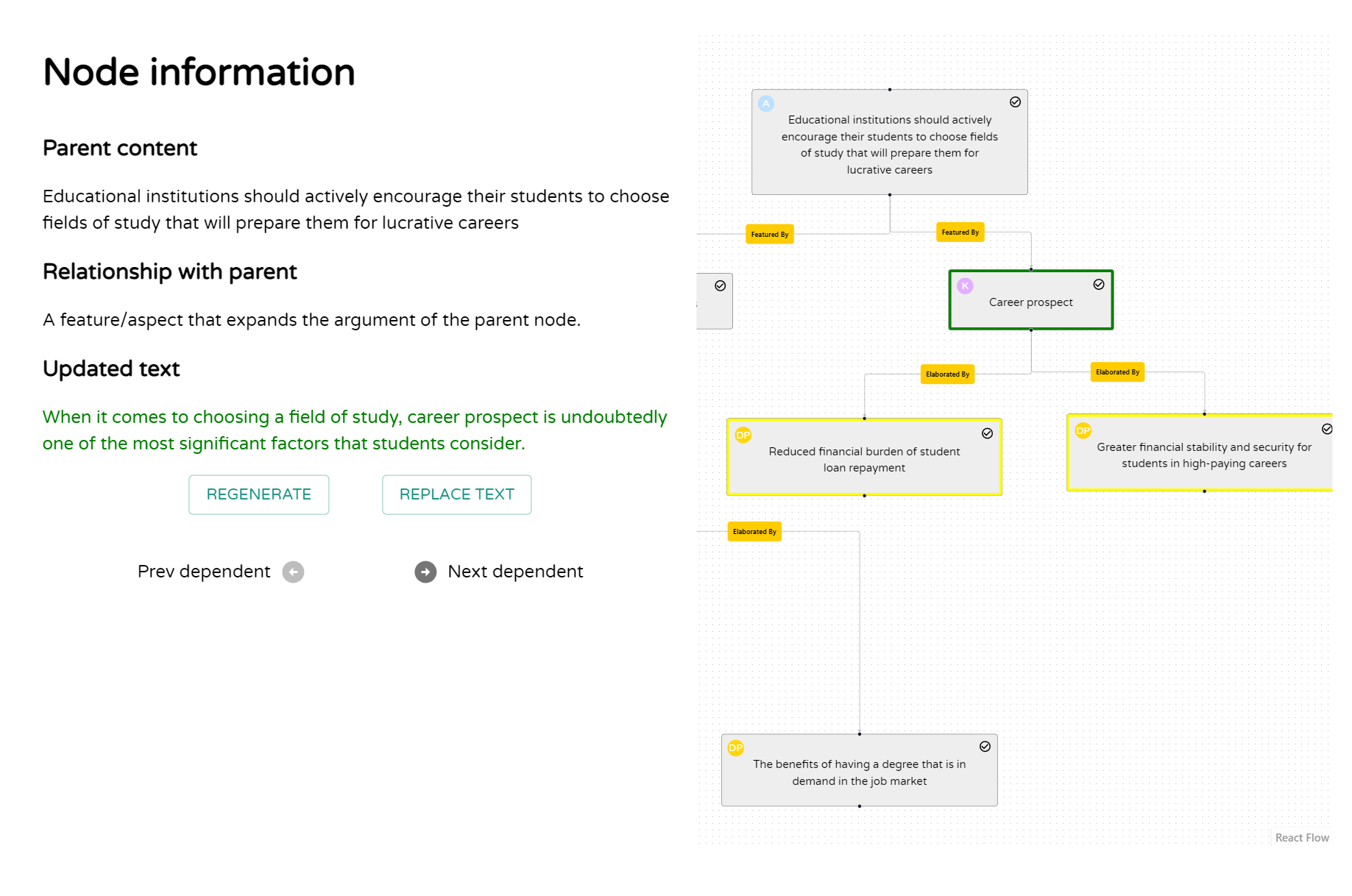}
  \caption{When the content of a node has changed, \visar prompts users to update all of its dependent nodes recursively.}
 
  \label{fig:update_modal}
  \vspace{-3mm}
\end{figure}

\subsubsection{Argumentative sparks}

As highlighted in DG3, \visar aims to provide scaffolding support for users to enhance the persuasiveness of their outlined argumentation. According to argumentation theory \cite{blair2011groundwork}, effective persuasive writing should comprehensively address counterarguments that could potentially weaken the main argument. Additionally, the writing should avoid logical fallacies and incorporate a diverse range of supporting evidence to boost credibility. Building on these points, we have provided \visar's underlying LLM with a set of few-shot examples that enable \visar to generate contextually relevant counterarguments, logical fallacies, and supporting evidence related to the current selected draft sentence or bullet point (see details in Section \ref{sec:spark_generation}).

\paragraph{Prompts for counterarguments}

Writers can explore potential counterarguments for a paragraph or discussion point by clicking on the corresponding text block, selecting \textit{Argumentative sparks}, and then \textit{Counter arguments} on the float menu. \visar will display a list of counterargument points generated by the LLM. Writers can choose multiple counterarguments to include in their outline. By clicking the \textit{Review and sketch} button (Figure \ref{fig:argumentative_flow}-3), they can add the selected counterarguments to the visual outline and review the current outline (Figure \ref{fig:elaboration_flow}-4). Finally, when writers clicking the \textit{Generate} button, \visar will produce a draft that describes the counterargument points in detail and insert it into the draft after the corresponding text block.

\paragraph{Prompts for logical fallacies} 

Similarly, writers can use \visar to identify potential logical fallacies of a discussion point in their outline. To do that, they click on the text block, select \textit{Argumentative sparks}, and then \textit{Logical weaknesses} on the float menu (Figure \ref{fig:argumentative_flow}-4). \visar will suggest a list of logical fallacies generated by the LLM that may apply to the selected discussion point. For each suggestion, \visar also displays an explanation of how this fallacy may apply to the paragraph. Writers can select the fallacies they want to address and, by clicking the \textit{Address weaknesses} button, \visar will present possible revision options for fixing the selected fallacies. Writers can click the replace (\faSync*) button to update the draft accordingly.

\begin{figure*}[htp]
  \centering
  \includegraphics[width=\linewidth]{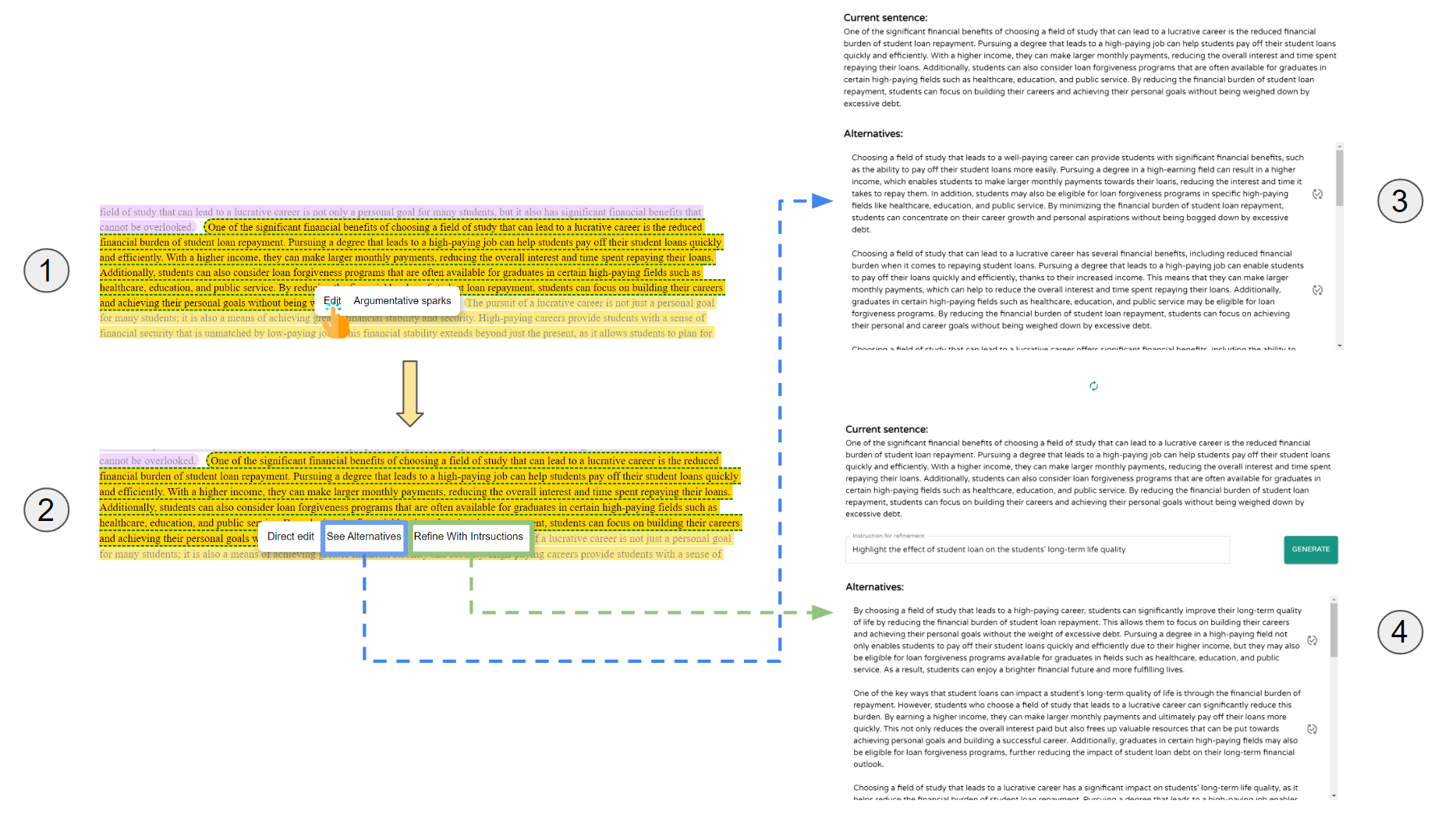}
  \caption{The interaction flow of text editing: (1) Users can edit the draft by clicking the corresponding text block; (2) Users can select different editing strategies; (3) Users can ask \visar to generate alternative drafts; (4) Users can provide \visar with flexible natural language instructions revising a specific part of the draft content.}
 
  \label{fig:edit_flow}
  \vspace{-3mm}
\end{figure*}


\paragraph{Prompts for supporting evidence}

\visar can suggest supporting evidence for a discussion point by adopting Aristotle's four persuasive strategies \cite{rapp2002aristotle}: \textit{ethos} (persuade by sharing professional experience), \textit{pathos} (persuade by arousing audience's emotions), \textit{logos} (persuade by indicating facts or logical reasoning), and \textit{example} (persuade by presenting concrete example). Details of suggestion generation are discussed in Section \ref{sec:spark_generation}. Writers can explore the available types of supporting evidence by clicking on the corresponding text block and then selecting \textit{Argumentative sparks} and \textit{Supporting evidence}. \visar will present users with potential types of evidence, accompanied by descriptions in the context of the selected discussion point. Writers can add the selected evidence types and review the updated visual outline by clicking the \textit{Review and sketch} button (Figure \ref{fig:argumentative_flow}-5). Upon confirming the outline, they can click \textit{Generate} to draft prototype paragraphs that implement the selected evidence types.\looseness=-1

\subsubsection{Rapid draft prototyping}
\label{sec:draft_prototyping}

As emphasized in DG4, \visar should support users in quickly generating drafts, allowing them to reflect upon, evaluate, and iteratively improve their current plan. \visar offers flexible draft generation timing, helps maintain consistency among interdependent nodes, and assists in updating the draft.

\paragraph{Supporting different timing of prototyping}

Writers may have varying preferences on the timing of draft generation during the planning process. Some writers might prefer to view the draft of an individual component immediately upon its addition to the outline, thereby enabling a direct assessment of its suitability. Conversely, others may prefer to concentrate on constructing the outline and initiate the draft generation process only after completion of the entire outline. To accommodate this, \visar supports flexible timing of draft generation. With the \textit{Lazy update} mode off (Figure \ref{fig:VISAR system}-D), users see the updated draft immediately after modifying the visual outline. When the mode is on, users can edit the outline and click the \textit{Generate text} (\faPencil*) button (Figure \ref{fig:VISAR system}-F) to update drafts for all components at once.

\paragraph{Maintaining consistency among interdependent components}

The writing components could be highly interdependent, which complicates the maintenance of consistency. For example, if a writing component has child dependents (e.g., counter argument, supporting evidence), modifying the component could cause inconsistency between its new content and the original children. To address this, \visar allows writers to \textit{recursively} update dependent children components. In the update window (Figure \ref{fig:update_modal}), writers can navigate between affected dependent components using \faIcon{arrow-alt-circle-left} and \faIcon{arrow-alt-circle-right} buttons. Upon moving to a dependent component to be updated, users can select a new topic from a recommended list and generate the updated draft.

\paragraph{Flexible editing strategies on the draft}

\visar allows writers to edit and update drafts using three strategies (Figure \ref{fig:edit_flow}). They can manually edit a block of text by clicking the \textit{Direct Edit} option. In addition, users can request \visar to propose alternatives using the \textit{See Alternative} button. Writers can replace the old component with an alternative (Figure \ref{fig:edit_flow}-3) by clicking the \textit{Replace} (\faSync*) option. Lastly, the \textit{Refine With Instruction} option allows users to interact with \visar in a ChatGPT-like conversational interface (Figure \ref{fig:edit_flow}-4), where writers can provide flexible instructions to refine the selected component. Writers can engage in multiple iterations of edits to refine the draft to their satisfaction.

\subsection{Implementation}

\subsubsection{Web application}


The interactive Web app of \visar is implemented in React. Specifically, the text editor is constructed with Lexical\footnote{\url{https://lexical.dev/}}, which models and organizes the editor's content into a hierarchical structure comprising nodes. We designed custom node classes for each type of writing component supported by \visar (i.e., key aspects, discussion points, counterarguments, and evidence). The back-end server uses the Flask framework\footnote{\url{https://flask.palletsprojects.com/en/2.2.x/}} for communication with the LLM and accessing a MongoDB\footnote{\url{https://www.mongodb.com/}} database.

\subsubsection{Writing goal recommendation}

\visar used OpenAI's API (GPT-3.5-turbo\footnote{\url{https://platform.openai.com/docs/models/gpt-3-5}}) to recommend writing goals. The detailed prompt templates are shown in the Appendix. Specifically, the model is first prompted to play the role of ``\textit{a helpful writing assistant that aims to help writers come up with high level aspects or topics that they can think of to support their argument}''. As users select arguments and click the \textit{Elaborate} button to obtain inspiration for key aspects, we prompt the model with "\textit{Please list key aspects that are worth discussing to support the argument: [selected argument]}." We also provide the model with several examples in the prompt, which inform it of the desired output format (i.e., a numbered list). 

Similarly, to elicit \textit{discussion points} regarding a key aspect of an argument, we direct the model to adopt the role of ``\textit{a helpful writing assistant that aims to come up with pertinent discussion points based on a specified aspect to reinforce the given argument}''. We supply the model with several examples and prompt: ``\textit{List key discussion points worth including in the discussion to support argument [selected argument] from the aspect of [selected aspect]}''. This prompt is executed for each selected aspect. In practice, we discovered that employing a ``chain of thought'' strategy (initially considering aspects and subsequently generating discussion points based on the selected aspects) allows the model to produce more detailed and relevant discussion points. This observation confirms the findings of previous studies~\cite{wei2022chain, saparov2022language} that prompting LLMs to undertake explicit intermediate steps enhances their problem-solving and reasoning capabilities.


\subsubsection{Argumentative spark generation}
\label{sec:spark_generation}

\visar leverages the GPT 3.5 model to create argumentative sparks\footnote{See prompt templates in Appendix}. Instead of relying solely on the model's zero-shot capabilities, we supply it with several examples of each type of argumentative spark  (counterargument, logical fallacy, and supporting evidence) in the prompts. These examples are based on proven methods for identifying counterarguments or taxonomies of logical fallacies and supporting evidence from argumentation theory, ensuring that the generated sparks are consistent with established principles of argumentation. We found that, when we provide the model with a few examples, it can proficiently consider a wide range of categories, even if the model is not explicitly shown all of these categories. The detailed example and prompts can be found in the Appendix.

\paragraph{Counterargument generation}

Prior research in argumentation theory \cite{blair2011groundwork, van2013fundamentals} suggests various strategies for opposing an argument, such as attacking the premise, warrant and backing component of the argument, or raising an alternative explanation for a phenomenon.  We familiarize the model with the strategies summarized by Blair~\cite{blair2011groundwork} by providing it with several examples mentioned in the Blair's work. Specifically, we inform the model about the applicable counterargument types and their application to particular arguments in the examples. When writers click the \textit{Counterargument} button on a selected argument, we prompt the model to act as a ``\textit{helpful writing assistant specializing in argumentative essay tutoring and generating counterarguments for the given statement}'', then enter the following prompt: ``\textit{Here are several examples of how to raise counterarguments for an argument: [examples]. Please list potential counterarguments that can challenge the argument [selected argument]}''.

\paragraph{Logical fallacy generation}

To identify logical fallacies in a discussion point, we use the taxonomy of logical fallacies summarized by argumentation theories \cite{blair2011groundwork, eemeren1987handbook} and guide the model to identify which fallacy types may apply. During inference, we first prompt the model to play a role as ``\textit{a helpful writing assistant focusing on argumentative essay tutoring and trying to suggest logical weaknesses in the given statement}'', and then input the prompt: ``\textit{Here are several examples of demonstrating how to find logical weaknesses for an argumentation: [examples]. Please list potential logical weaknesses in the argument [selected argument]}''. 

\paragraph{Supporting evidence generation}

We introduce the model to to Aristotle's four persuasive strategies~\cite{rapp2002aristotle} (\textit{ethos, pathos logos, and example}) for generating supporting evidence for a selected argument. Specifically, we first ask the model to play a role as ``\textit{a helpful writing assistant focusing on argumentative essay tutoring and trying to suggest supporting evidence types in the given statement}''. During inference, we prompt the model: ``\textit{Please list potential supporting evidences that can back the argument: [selected argument]. You can think from the following aspects: sharing professional experience (ethos), arousing audience's emotion (pathos), providing facts and strict logical reasoning (logos) and presenting concrete practical examples (example)}''. 

\subsubsection{Draft generation}

We utilize the model to generate prototype drafts that implement writers' argumentation plans. The prompt templates for creating each specific writing component, as well as for revision and refinement, are detailed in the Appendix \crhighlight{(Table \ref{table:prompt_templates})}. When a writing component has a parent node in the plan, we include the parent's content as context to improve coherence between the two elements. 



\section{User Study}

We conducted a lab user study\footnote{The study protocol was reviewed and approved by the IRB at our institution.} with 12 participants to evaluate the usability, effectiveness, and usefulness of \visar. The study aims to answer the following questions:


\begin{itemize}
    \item \textbf{RQ1:} Can users successfully plan argumentative writing using \visar? 
    \item \textbf{RQ2:} How useful is \visar in facilitating the ideation, organization, and revision of users' writing planning?
    \item \textbf{RQ3:} What challenges do users encounter when using \visar to plan their writing?
\end{itemize}

\subsection{Participants}

We recruited 12 participants from a private R1 university in the United States. The participant group comprised 1 sophomore, 3 juniors, 4 seniors, and 4 graduate students. 4 participants identified themselves as intermediate writers (having some experience but still honing their skills), 7 as advanced writers (with significant experience and confidence in their abilities), and 1 as an expert writer (having extensive experience and a high level of skill). We anticipated that most, if not all, of our participants would be proficient writers. It is important to note that the average incoming freshman at this institution scores in the 96th percentile in the Evidence-Based Reading and Writing (EBRW) section of the SAT test, and all students complete two semesters of Writing \& Rhetoric in their first year. All participants were proficient in English, and 8 were native English speakers. Each participant received a \$30 USD compensation for their time.


Regarding their experience with AI-supported writing tools, 8 participants had used AI tools for grammar and spelling assistance, 7 for writing prompt or idea suggestions, 5 for refining writing styles or tones, 6 for polishing content or language, and 1 for plagiarism detection. In terms of their familiarity with writing planning methods, 2 were familiar with mind maps and 11 were familiar with outline sketches. All 12 participants had previously used generative AI tools (such as ChatGPT) to support their work in various scenarios, including brainstorming outline and ideas, refining the language, summarizing text, and generating code snippets.

When asked about the challenges they commonly face in planning argumentative writing, 7 participants reported having previously experienced difficulties in generating ideas, 9 in organizing ideas and information logically, 3 in developing counterarguments, and 4 in making their arguments more persuasive.

\subsection{Study Design}

Each study session, lasting approximately 90 minutes, was conducted in-person at a usability lab. The study used a within-subjects design, where the participants experienced three conditions during three 25-minute argumentative writing planning sessions. The study procedure is detailed in Section~\ref{sec:study_procedure}. 

\subsubsection{Tasks}
\label{sec:study_tasks}
To design a representative domain-general task for argumentative writing, we adopted the Issue Essay task from the Graduate Record Examinations (GRE)\footnote{\url{https://www.ets.org/gre/test-takers/general-test/prepare/content/analytical-writing/issue.html}} for our study. This task necessitates that writers create an argumentative essay exhibiting critical thinking while clearly expressing their thoughts on a given topic that can be approached from various perspectives and applied to multiple situations and conditions. In our study, we selected three topics from the GRE's issue writing sample pool\footnote{\url{http://words.gregmat.com/greessay.html}}. Our selection criteria stipulated that the topic should not require expert knowledge and must have an appropriate scope, allowing writers to generate specific points from different aspects within a limited time frame. The selected topics and the corresponding task descriptions are provided in Appendix. Users encountered a different topic in each writing session, during which they were asked to use the system corresponding to the current condition to create an outline for the argumentative essay and draft as much of the essay as possible. The order in which the participants addressed the three writing topics was randomized.


\subsubsection{Conditions}

The study incorporated the following three conditions:

\begin{itemize}
    \item \textbf{Baseline}: A plain text editor without AI support.
    \item \textbf{GPT Playground}: A text editor resembling OpenAI Playground\footnote{\url{https://platform.openai.com/playground}} that allowed users to input prompts and select text for the LLM (GPT-3.5-turbo was used, the same as \visar) to generate a response. The model's output was appended to the participants' prompt in place. 
    \item \textbf{\visar}: A full version of \visar featuring visual programming, step-by-step ideation support, argumentative sparks, and rapid draft prototyping capabilities.
\end{itemize}

We used a Latin square experimental design \cite{ryan2007modern} to ensure a balanced sequence of task topics and conditions for each study session.

\subsubsection{Study procedure}
\label{sec:study_procedure}

At the beginning of each study session, the experimenters collected informed consent and demographic information from the participants. The study coordinators then provided a high-level introduction of the writing tasks (as described in Section \ref{sec:study_tasks}). Each participant completed three writing sessions, each session comprising a 5-7 minute tutorial and a 20-minute writing planning task. The tutorial explained the key features of the tool associated with the current condition.  The task sessions were independent of each other, which means that the output created in each condition was not carried over to subsequent conditions. After the participants completed the three writing sessions, they completed a post-study questionnaire, \crhighlight{which mainly assessed usability, usefulness, and enjoyment of \visar system}. The study ended with a 10-minute semi-structured interview, during which the experimenter asked questions to understand the rationale behind the participants' observed behaviors, tool usage, and reflections on their experiences with the tools. \crhighlight{The purpose of the semi-structured interview was to explore participants' overall experiences, and to elicit their specific feedback on VISAR features, final draft quality, improvement suggestions, and potential benefits from the system}.





\subsection{Result}
\subsubsection{Post-study Questionnaire}
\begin{figure*}[!t]
    \centering
    \includegraphics[width=\linewidth]{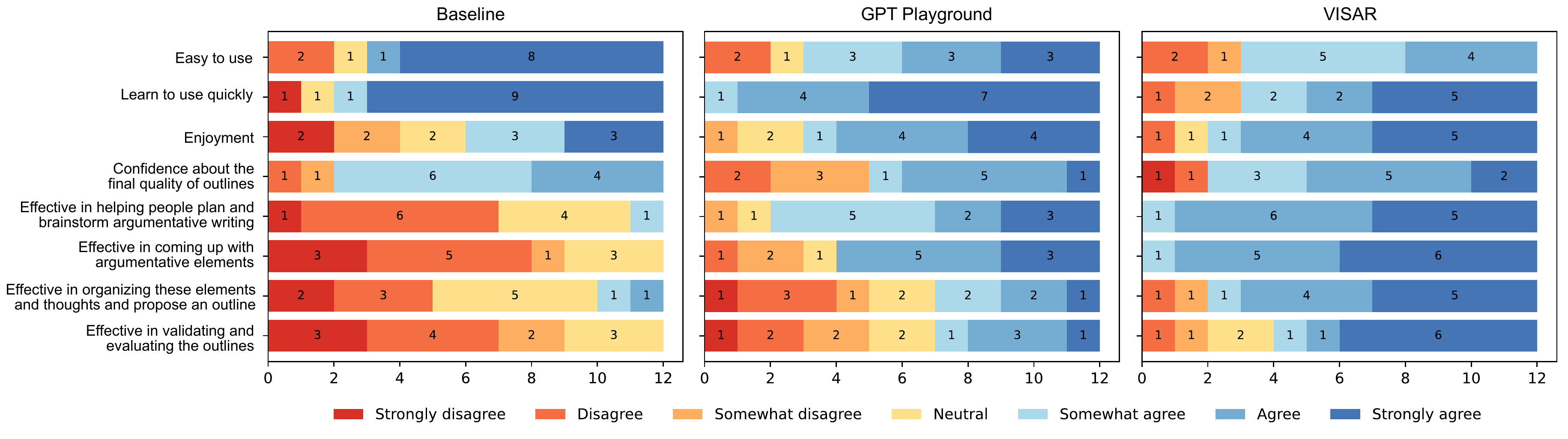}
    \caption{Results from the post-study questionnaire comparing user ratings of three conditions}
    \label{fig:results:questionnaire}
\end{figure*}

Figure~\ref{fig:results:questionnaire} summarizes the results of the post-study questionnaire of 12 participants. In general, the participants found that \visar was more effective in helping them generate argumentative elements, organize these elements into outlines, and validate the outlines compared to the GPT Playground and baseline conditions. The participants also expressed confidence in the quality of outlines produced using \visar and considered the system easy to learn and enjoyable to use.

Although \visar did not score as high as the other two conditions on ``easy to use'' and ``learn to use quickly'' due to the many new features added, our participants still considered it generally easy to use.


\subsubsection{Post-study interview}
\label{sec:interview_result}
Following established open-coding protocols \cite{braun2006using, lazar2017research}, two study team members first performed a round of independent qualitative coding of interview transcripts. Afterward, they had a discussion to achieve agreement and created a consolidated codebook. Using this codebook, a thematic analysis was performed to uncover emerging topics from the interview. The entire research team collectively reviewed the coding outcomes to identify high-level themes. We report the following key findings:

\paragraph{\textit{KF1: Assistance for hierarchical ideation aids users in developing structured outlines.}}

The participants found the hierarchical goal recommendations provided by \visar useful to help them explore and develop structured outlines. The participants found that \visar could ``\textit{give us (them) more points than we (they) would have thought about}'' (P3), allowing them to ``\textit{realize the missing thinking aspects and efficiently explore extensive possible directions by a few clicking}'' (P9). In addition, some participants found that the ideation assistance process of \visar aligned well with their natural thinking process. For example, P7 said ``\textit{it was kind like what I did in my own writing, creating opening sentence and then thinking of concrete aspects and examples to support my argument step by step. It was like compile whatever the thing was and gave me some options at each thinking step, except through guided by machines}''. Confronted with this familiar AI-guided ideation process, the participants could ``\textit{reflect and reassure their own thinking}'' (P2), ``\textit{articulate their thoughts}'' (P8) and ``\textit{better understand their own arguments}'' (P8).

However, some participants worried that this ideation assistance could ``\textit{deprive our (their) control of the writing directions}'' (P7) and make them ``\textit{less creative and initiative in thinking about the writing}'' (P9). Besides, this hierarchical assistance is limited to help on the non-linear ideation part of the planning. For example, P4 found that ``\textit{this assistance did not help me (her) understand how I (she) could link or compare two aspects included in my (her) outline}''.



\paragraph{KF2: The synchronized text and visual views provide complementary assistance for writing planning}

All participants reported that text and visual editors helped their writing planning in different ways, and the color coding of text and node helped them easily correspond the two modalities. The visual editor helped them understand the clear logical structure of their outline and draft, as P12 said, ``\textit{it is pretty useful because when we (they) want to write like a good argumentative essay, we (they) wanted to be logically clear}''. Also, the visual editor eased their work to edit and organize their outline. For example, P11 said ``\textit{we may keep adjusting the outline like making your argument followed by an example or led by another counter argument, without this tool (visual editor), it will be pretty hard for us to logically put plenty of arguments together}''. In addition, visual editor could also help them quickly ``\textit{navigate to a concrete part of outline}'' (P3) and ``\textit{find the missing part}'' (P7). The participants also found the visual editor useful for supporting their ``\textit{two-stage cognitive process including planning and reviewing}'' (P1), where they first ``\textit{draft and think how things can go hierarchically  and then see whether I (they) can add more nodes or move a node that should be fit better under another category}'' (P4).

On the other hand, the participants found the text editor to be helpful and easy to use for reviewing and modifying the concrete draft of a discussion point after deciding on its position in the outline. For example, P2 said ``\textit{I feel like this argument should go off this branch and then it would just like make sense to do it from the tree part versus doing it from the editor and then I went on the (text) editor once I like made that first initial branch and saw okay this is where it's gone and then I was able to edit from there}''. Compared to the visual editor, the text editor provided the participants with specific contexts so that they can better ``\textit{validate the cohesiveness of their arguments}'' (P5). Besides, some participants preferred to use the text editor for outline sketching because ``\textit{it aligns with what I (they) normally did as drafting an essay}'' (P7).

\paragraph{KF3: Argumentative sparks help users strengthen their writing}

The participants found that the argumentative spark features help them enhance the persuasiveness of their writing. Specifically, these argumentative sparks help them identify opportunities to improve a particular argument. For example, P6 said ``\textit{It was really cool to see like you can generate counter arguments for this section, because I feel like I wasn't focusing enough on counter arguments. So that is helpful to look at like potential logical fallacies}''. Besides, the participants reported that the recommendations of counterarguments and evidences were helpful for them to improve the objectivity of their draft and articulate their thoughts. For example, P8 said ``\textit{I feel like it's very helpful, especially in like an argumentative writing setting, because it remains relatively objective when it talks about things. Yeah, especially useful when you have an idea in mind that you don't really know how to articulate it}''.

However, the participants also found that the argumentative sparks were sometimes not sufficiently concrete. For example, P3 said ``\textit{I think the only issue I I was coming up, I asked for supporting evidence for one of my points. And instead of giving me like supporting evidence, the only thing that the the bubble button said was one possible evidence showing that there is scientific research to prove that this is right. So it didn't really tell me what the scientific research was if that makes sense. So it didn't give me the point, it just won't be worth a point.}''.

\paragraph{KF4: Draft prototyping helps users improve plans but it cannot be used directly in the final product}

Most of the participants thought that the rapid prototype feature could help them understand and improve their planning within a concrete context, although the generated draft tended to suffer from issues such as robotic tones, repeating content, poor transition, and monotonous language style across different generations. An important benefit mentioned by many participants was that the draft prototyping allowed them to preview what the implementations of their plans \textit{could} look like as they developed the plan. This helped the participants go back to the planning stage to expand their thought and come up with more specific points, which was otherwise difficult to achieve at the high-level planning stage. For example, P5 said ``\textit{I initially started with like really basic things. Then I expanded them a little bit more when I saw those generations, because it reminded me of some more specific responses and more detailed evidences that I didn't think of when drafting the outline. I think it was pretty key, pretty important}''. Besides, the participants thought that this feature could dramatically save their effort and time in creating a prototype draft to validate their ideas and support further revision. For example, P11 said ``\textit{Overall, while I may be picky about how things are phrased, I like being able to directly edit on the rough draft like it, based on how I think ultimately is best. That may be more helpful and saves time to write like three sentences versus doing it on my own. It takes me to think directly about how to phrase those three sentences and improve my arguments}''. The automatic prototyping also acts as a ``milestone'' that provided the participants with a sense of achievement, as P6 commented, ``\textit{It kind of made you feel like you made a lot of press. you're like, I only have like four nodes, but here's a whole starting essay}''.

\paragraph{Challenges, suggestions and additional application scenarios}

The participants faced usability challenges related to interaction and display. For instance, P5 and P8 reported difficulties connecting two nodes in the visual editor due to the small size of the connection handles. Additionally, the initial position of the AI-generated nodes and the space between them were sometimes problematic (P3, P5) due to an implementation bug, particularly when participants incorporated a large number of AI-recommended options into their plan simultaneously. Moreover, the participants observed that the order of the paragraphs in the draft sometime fails to update when they adjusted the order of the corresponding visual nodes (P5, P10).

In addition to interaction challenges, the participants noted that \visar had a steeper learning curve compared to the systems under the other two conditions because they had to learn how to use the new features. In some cases, the use of \visar may also increase their cognitive load, as they had to ``read the long generated draft to make sure it aligns with my (their) intent'' (P3).\looseness=-1

The participants offered valuable recommendations to improve and expand the current system. P9 proposed adding a feature to generate visual mind maps from users' existing drafts, assisting users in analyzing their draft's logical structure. P1 suggested extending the use of argumentative sparks to entire outlines (in addition to individual discussion points) to improve overall persuasiveness. P3 recommended that \visar include citations for the suggested supporting evidence, which would require technological advances in LLMs. Lastly, P6 believed that \visar should offer more intuitive controls, such as allowing users to drag paragraphs to reorder them and update the visual mind map accordingly.

Regarding potential application scenarios for \visar beyond argumentative writing planning, many participants suggested that \visar could assist with brainstorming and organizing ideas for lighter writing tasks that require quick and logical drafts. Additionally, P10 mentioned that \visar could be used during the revision stage to facilitate critical evaluations of drafts by providing argumentative sparks. However, some participants expressed caution about using \visar's output directly in writing for more serious purposes, such as semester papers (P1) or personal statements (P9), due to ethical concerns about plagiarism and authorship (P6), and doubts about \visar's ability to accurately represent their personal experiences and knowledge (P8).

\begin{figure}[!t]
    \centering
    \includegraphics[width=\linewidth]{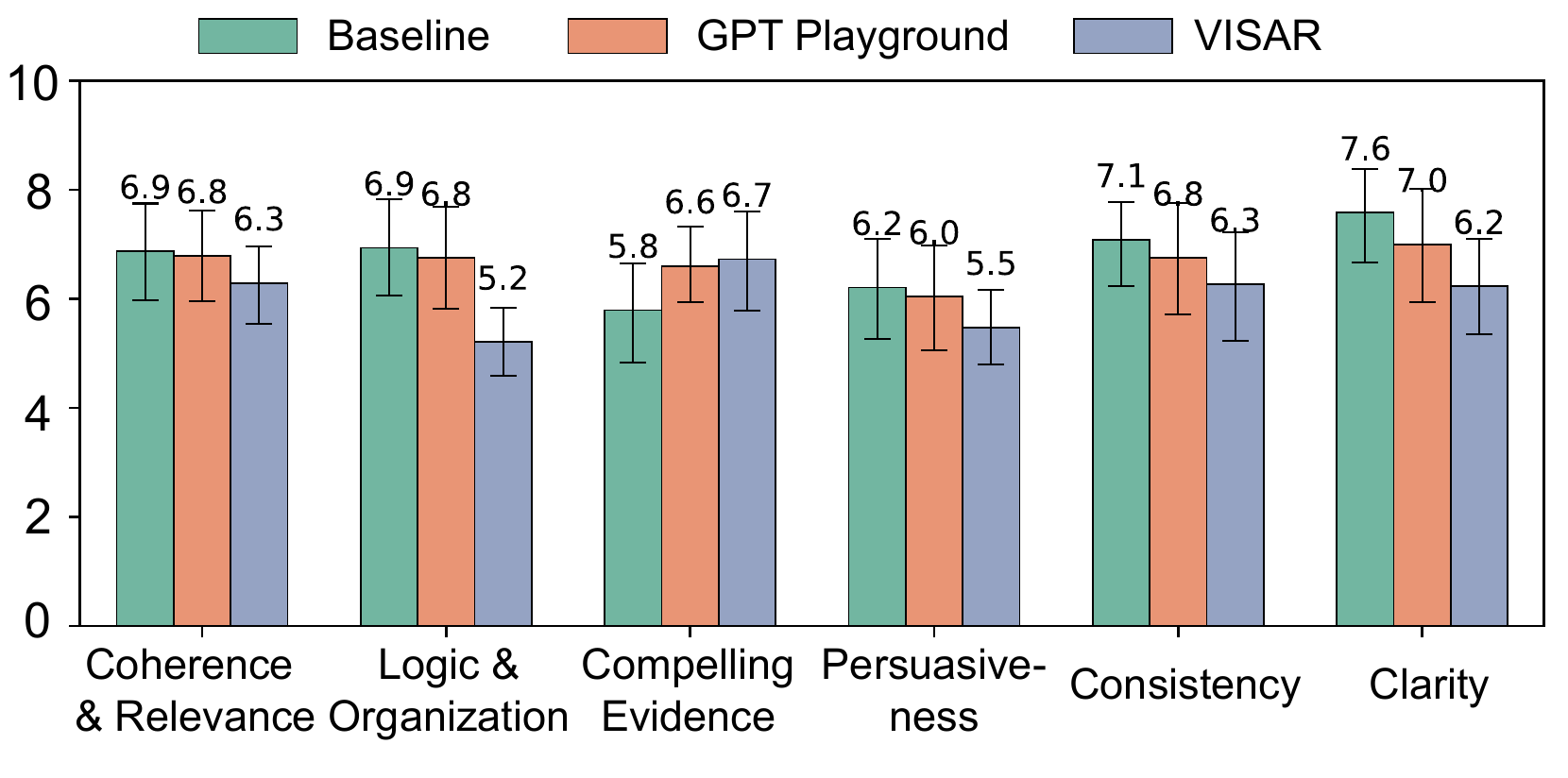}
    \caption{Average expert scores on each evaluative dimension across varying conditions.}
    \label{fig:results:evaluation}
\end{figure}

\subsubsection{External Evaluation of Participants' Drafts}

\crhighlight{While our main focus was on the prewriting and planning phases of argumentative writing, we also investigated the quality and characteristics of the drafts generated by our participants. It's important to note that participants were not required to complete a full draft on their assigned topic in each condition; their primary task was to form a robust outline (as discussed in Section \ref{sec:study_tasks}). Yet, evaluating and comparing the quality of these provisional drafts across different conditions can offer insights into the advantages and potential problems of transitioning from plans to drafts with each interface}.



\crhighlight{We have two experts in argumentative writing evaluate the quality of each draft collected in our studies}. Both experts were Writing \& Rhetoric instructors at a private U.S. university with extensive professional experience in practicing, teaching, and assessing argumentative writing. 

The expert evaluators were given a set of assessment criteria adapted from the official GRE website\footnote{\url{https://www.ets.org/gre/test-takers/general-test/prepare/content/analytical-writing/scoring.html\#accordion-06cf390d3c-item-60c4f150ea}}, Re\textsuperscript{3} \cite{yang2022re3} and Adaptive Learning system \cite{wambsganss2020adaptive}, where they rated each submission on the following six dimensions on a scale of 1 to 10. They also provided short feedback on the strengths and weaknesses of each outline. \crhighlight{During the evaluation process, the evaluators were unaware of the conditions under which the drafts were produced. The evaluative dimensions were shown below}:

\begin{enumerate}
    \item Coherence and Relevance: identify aspects of the argument that are pertinent to the assigned task and examine them with insightful analysis.
    \item Logical organization: develop ideas and arrange them in a logical sequence that supports the overall argument.
    \item Compelling and thorough evidence: present strong, comprehensive evidence that substantiates the arguments made.
    \item Persuasiveness: evaluate the effectiveness of the argument in persuading the target audience.
    \item Consistency: ensure that the argument is factually consistent and free of contradictory details.
    \item Clarity: assess whether the argument is easily understandable and presented with clear language and structure.
\end{enumerate}


\crhighlight{\subsubsection{Quantitative Results} A summary of the ratings on each evaluative dimension across conditions is shown in Figure~\ref{fig:results:evaluation}. The scores indicated that though drafts from \visar condition have passing quality, they still fall short from excellence.} 

\crhighlight{In particular, a one-way repeated measures ANOVA was conducted to investigate the influence of three conditions on the six measures. Our findings  revealed a significant main effect on the score of Logic \& Organization from the use of \visar ($F_{2, 46} = 6.36, p < .01$). Post-hoc pairwise t-tests revealed that the score on Logic \& Organization of \visar ($M=5.2$) is significantly lower than both the baseline ($M= 6.9, p=.02$) and GPT playground ($M= 6.8, p<.01$). The difference between the baseline and GPT playground was not found to be statistically significant. We found no significant difference among ratings on all other dimensions among the conditions.}

\crhighlight{\subsubsection{Qualitative Results} To understand the characteristics of drafts produced in each condition, two researchers conducted qualitative coding~\cite{mcalister2017qualitative} on evaluators' open-ended comments on the drafts. Specifically, each researcher independently annotated expert comments using either \textit{vivo} codes or self-defined codes. Then they sat together to discuss, came to a consensus on changes to the codebook, and relabeled comments independently using the newly formed unified code book. The Cohen’s kappa score between the coding results from the two researchers in the second round was 0.66, indicating substantial agreement~\cite{landis1977measurement}.}


\crhighlight{In terms of strengths and weaknesses, evaluators found that user-written drafts from baseline condition tends to have clear and cogent main arguments. They usually contained logical and coherent argumentation structures and ``\textit{there is a clear clue how the draft was expanded from and follows the outline}''. However, those drafts tends to lack details such as specific examples, supporting evidences or counterarguments, which could cause ``\textit{the argumentation stays on a surface level}''.}\looseness=-1

\crhighlight{In the case of the GPT Playground condition, evaluators observed that these drafts were filled with a large number of details that gave the arguments an appearance of solidity. However, they also noted that such drafts could be poorly structured, having an unclear main argument (or even contradictory arguments) and an unsatisfactory flow of reasoning. In addition, some examples and evidence provided in the drafts did not adequately support the immediate arguments, causing the argumentation to veer off topic and become disjointed in certain instances. This could be partly attributed to participants frequently invoking ``auto completion'' to understand how an outline item could be expanded. Due to time constraints, they often exerted minimal effort to modify the generated content, reorganize the flow, or remove unsuitable completions. }

\crhighlight{Finally, evaluators found that the drafts from the \visar condition were abundant in specific content drawn from a variety of perspectives and at different depth levels. However, they noted these drafts often suffered from issues like repetitive or self-conflicting content, and unclear transitions and relationships between paragraphs and sentences. For example, evaluator reported issues like ``\textit{The product felt like a series of mini-essays collated together without any transitions or framing. This makes it difficult to remember the writer's main argument and affects the essay's overall effectiveness}''. Nonetheless, evaluators thought that some of these drafts ``\textit{have great potential of becoming a compelling essay if authors organize the content in a better way}''. }

\crhighlight{These qualitative findings suggest that while a generative AI model can aid in creating draft and enable writers to anticipate how the outlines may unfold, humans still need to supervise the progression from plan to draft. It is essential to ensure that transitions and logical relationships between writing elements are valid and clear. A major limitation of the current \visar system is that it does not sufficiently consider the high-level context of the article when generating content. Future AI-augmented writing systems could incorporate external memory repositories\footnote{For example, we can incorporate external memory using LLM application framework like LangChain: \href{https://python.langchain.com/docs/modules/memory/}{https://python.langchain.com/docs/modules/memory/}} (e.g. vector database) or more effective prompting strategies (e.g. self-reflection~\cite{shinn2023reflexion} or tree-of-thoughts~\cite{yao2023tree}) to enhance the coherence and relevance among the generated content.}

\section{Discussion}





\subsection{Interaction Modalities and Strategies for Collaborating with LLMs}

The ``default'' modality for interacting with LLMs is through natural language conversations (e.g., ChatGPT). However, this mode of interaction poses challenges to human-AI collaboration and limits the potential applications of LLMs. From the perspective of the AI model, models encounter difficulties in accessing the specific task context and user behavior history, which are often disconnected from the conversational context, making it difficult to understand the implicit task specifications and user intent \cite{borji2023categorical,li_sovite:_2020,li_pumice:_2019}. This can result in inaccurate or irrelevant responses to user requests. 

On the user side, unlike using cognitively friendly tools such as visual diagramming, users may have trouble expressing their intent or needs in textual form~\cite{whitley1997visual} because the interaction is ``too open'' without many indicators for the possible actions at the current state, the system's capabilities, and the most useful types of input to provide for the system. Furthermore, users must transition to a conversational environment to interact with LLMs, potentially imposing additional cognitive load. These challenges can be particularly significant in complex real-life scenarios where tasks may have multiple sub-parts and users utilize different tools or modules to address them.

\visar addresses these challenges by simultaneously enabling direct manipulation and conversational agent interfaces across different modalities, allowing users to interact with AI models in a modality that best supports their current task in a unified workspace~\cite{shneiderman_1997_direct}. For instance, direct manipulation, as an interaction modality, offers advantages such as intuitiveness, implicit representation of intent, awareness of the interaction context, and low cognitive load \cite{hutchins1985direct}. Therefore, we employ it as an interaction strategy for tasks where users might otherwise encounter difficulties in articulating their needs and corresponding context (e.g., argumentative sparks and hierarchical goal recommendation). Moreover, we use direct manipulation to facilitate user interaction with AI models in visual outlines, following the implications and recommendations of previous work~\cite{xia2022persua, young1995cantata}. Conversely, we employ conversational agent interfaces to enable users to explicitly convey and incrementally refine their requirements for draft generation (e.g., refining with instruction, addressing logical fallacies), where users need more flexibility and expressiveness that are difficult to provide in a direct manipulation environment. But in such scenarios, the availability of concrete drafts allow users to refer to them in their natural language instructions, alleviating the challenges in the lack of context when the user interacts with a conversational interface in earlier stages of the process.


\subsection{Generative AI for Rapid Prototyping in Creative Tasks}

Creative tasks often involve a non-linear process for generating new ideas and transforming them into tangible results. Based on the cognitive theory of creativity, the process includes key stages such as preparation, illumination, evaluation, and implementation \cite{kozbelt2010theories}. Individuals may cycle through these stages repeatedly, revisiting earlier phases to refine and iterate their ideas. Rapidly drafting prototypes throughout this process can be beneficial, as it allows individuals to reflect on their concepts, gain a clearer understanding of their alignment with the ultimate goal, and draw inspiration from the surrounding context. However, prototyping can be time consuming and labor intensive, as discussed earlier regarding argumentative writing.

Generative AI models have the potential to enhance the creative process with their ability to rapidly generate parallel prototypes. For example, \visar shows the feasibility and value of quickly producing drafts based on argumentative outlines. These drafts offer a tangible intermediate representation for both humans and models to understand the writing context, enabling models to assist individuals in evaluating their existing ideas, clarifying ambiguities in human-AI communication, and identifying potential gaps in the working scenario. The writers can create and organize different parallel argumentation plans using the visual outline feature, generate concrete drafts that implement these outlines with little effort, and compare the drafts in order to understand the strengths and weaknesses of different argumentation plans and iteratively improve them. 

However, the use of generative AI during the prototyping stage of creative projects may present challenges. One concern is that when humans do not create the prototypes themselves, they may need to invest extra effort to review and understand the generated content. As P3 noted, they \textit{had to read the lengthy draft to ensure it aligned with their original intent}. Additionally, machine-generated prototypes could unintentionally introduce bias or constrain users' ``blue sky'' creative thinking, highlighting research opportunities to better enable users to explore a wide range of potential ideas and integrate their own insights into the drafting process.

\subsection{Visual Programming for Bridging User Workflow and Model Thought Process}

Visual programming interfaces have proven effective in facilitating user workflow across various application domains (e.g. collaborative work~\cite{swenson1993visual, wang2015docuviz}, task planning~\cite{hayhoe2003visual}, interactive learning~\cite{broll2017visual}), helping to clarify relationships and dependencies between concepts and enabling users to understand and navigate complex information~\cite{myers1986visual}. Visual representations also allow users to see the big picture, identify patterns, connections, and gaps in their ideas, which enhance brainstorming, problem-solving, and decision-making.

\visar demonstrates the promise of integrating user workflow with LLM's thought process features within a visual programming interface. Users can create organized writing outlines by defining goals and their connections at different abstraction levels using visual representations of arguments, discussion points, and evidence, as well as the relationships among them. Simultaneously, by addressing and implementing these goals in a top-down sequence, the model can follow each logical step proposed by the user, progressively building an understanding of the current writing context. This enables the model to generate content that aligns with the abstraction levels of the user's thought process.

\subsection{Ethical Considerations for AI Writing Assistance}

While generative AI models have been found useful many subtasks of the writing process, designers of human-AI collaborative writing tools must carefully consider ethical concerns. First, models could introduce bias, stereotypes, and misinformation \cite{borji2023categorical, gero2022sparks}, resulting in factually flawed or prejudiced outcomes. Furthermore, models may unintentionally generate content that closely resembles or reproduces existing sources from their training data, raising concerns about plagiarism or copyright infringement, even if the user is not intentionally copying from existing works. Additionally, directly incorporating AI-generated content into one's writing could call into question the writer's intellectual integrity and authorship of the written product.

\visar addresses these potential ethical concerns by primarily using generative AI models to assist writers during the prewriting and planning stages, generating early prototypes only for sensemaking and comprehension purposes. The design of \visar emphasizes user control and autonomy, ensuring that the human writer takes the lead in key creative aspects of the writing process. \crhighlight{To better mitigate the biases and stereotypes in generated contents, future work can be done to combine LLM's capability of self-critique and reflection~\cite{chen2023teaching, shinn2023reflexion} with external strategies such as augmenting LLM prompts with relevant factual knowledge from credible sources like Wikipedia, red flagging potential bias using independent bias detection model~\cite{raza2022dbias}, and pre-emptively informing users about potential biases in model-generated contents.} \looseness=-1

To further address the ethical concerns associated with generative AI models, it is essential to establish guidelines and best practices for their use in creative work. This may include transparently revealing the use of AI-generated content, encouraging critical assessment of AI-generated suggestions, and implementing strategies to identify and mitigate biases in output. The new human-AI collaborative writing paradigm introduced by \visar opens up new challenges and opportunities regarding these aspects, which we would love to further investigate with the rest of the research community as future work.

\section{Limitation \& Future Work}


The current version of \visar has several technical constraints. First, the fixed prompt templates used for drafting, recommendation, and argumentative sparks generation could be sub-optimal, potentially yielding similar model responses despite distinct input contexts (e.g., analogous language styles across various generations). Second, as P7 pointed out, the system currently does not account for transitions between generated discussion points in prototype drafts, which may result in reduced cohesiveness in full drafts of prototypes. Third, the system supports only a limited range of argument and relation types, lacking support for certain specific argument components, such as warrants and qualifiers, and specific styles of argumentation, such as bottom-up deductive argumentation \cite{hitchcock2006arguing}. \crhighlight{Lastly, our system can only support top-down tree-structured argumentation so far, however, argumentation could have more complex structures such as directed graph. }

In addition to \visar's implementation limitations, the GPT 3.5 model used in \visar may also generate factual inaccuracies, inconsistent responses, or contradictory statements in the prototype drafts \cite{borji2023categorical}. This may expose users to confusing and misleading information, and may have adverse second-order effects on how users conceive of research and the practices and methods involved in finding data and evidence. In the design of \visar, we emphasize that the generated draft is only a ``prototype'' for facilitating iterative prewriting and planning processes, mitigating the adverse impact of this problem.

Regarding the study design, there are several potential threats to the validity of our findings. Our tasks used argumentative writing topics from the GRE test. Although these tasks are considered adequate representations of general argumentative writing tasks without requiring domain-specific knowledge, it would be valuable to further explore \visar's use in more diverse real-world writing tasks through a deployment study. Expanding the demographics of the participants in future deployment studies will also help assess the ecological validity of \visar.\looseness=-1

\crhighlight{It is also worth mentioning that the current \visar system is not specifically designed for any particular user group. As a result, the argumentative components it supports (such as node and relation types) are relatively generic.} In the future, we plan to explore the potential application of \visar in educational settings. \crhighlight{For example, we would like to explore how \visar can help K-12 students to learn} argumentative writing.  We will also investigate the tool's efficacy in providing personalized guidance, enhancing the structure and coherence of arguments, and fostering critical thinking skills. \crhighlight{In addition, we will employ argument mining based models~\cite{wachsmuth2017computational, lawrence2020argument} to automatically validate the logical soundness of arguments generated by LLMs and notify learners about any potential issues in those outputs. Lastly, we will explore how to accommodate additional argumentation structures other than tree-based one in the furture version of \visar.}\looseness=-1

Moreover, we plan to extend \visar to accommodate collaborative writing planning \cite{yim2017synchronous, wang2017users, wang2015docuviz}. Our research will focus on enhancing the system's functionality to support seamless collaboration among multiple users. We will develop shared visual programming spaces that allow users to jointly construct and refine their writing plans. Additionally, we aim to investigate the impact of these collaborative features on group argumentative writing quality and individual engagement. By incorporating effective collaborative elements into \visar, we seek to create a more engaging and interactive platform that promotes peer learning, diverse perspectives, and the co-creation of well-reasoned arguments.

\section{Conclusion}

This paper presented \visar, an AI-enabled tool designed to support the prewriting and planning stages of argumentative writing. \visar provides: (1) hierarchical writing goal recommendations to facilitate writers' ideation processes, (2) synchronized text and visual editors for easy planning and organization of outlines, (3) argumentative sparks that supply writers with supporting evidence, logical fallacies, and counterarguments, and (4) rapid draft prototyping, enabling writers to understand and improve their outlines by reflecting on draft implementations. A user study with 12 participants demonstrated that writers can effectively use \visar to create argumentative writing plans, and the system was perceived as useful, usable, and likable.

\begin{acks}
This work was supported in part by an AnalytiXIN Faculty Fellowship, an NVIDIA Academic Hardware Grant, a Google Cloud Research Credit Award, a Google Research Scholar Award, and an Asia Research Collaboration Grant from Notre Dame International. Any opinions, findings or recommendations expressed here are those of the authors and do not necessarily reflect views of the sponsors. We would like to thank John Behrens, Sherry Tongshuang Wu, Hua Shen, and Ting-Hao (Kenneth) Huang for useful discussions.
\end{acks}

\bibliographystyle{ACM-Reference-Format}
\bibliography{paper/maindraft}


\begin{thebibliography}{87}


\ifx \showCODEN    \undefined \def \showCODEN     #1{\unskip}     \fi
\ifx \showDOI      \undefined \def \showDOI       #1{#1}\fi
\ifx \showISBNx    \undefined \def \showISBNx     #1{\unskip}     \fi
\ifx \showISBNxiii \undefined \def \showISBNxiii  #1{\unskip}     \fi
\ifx \showISSN     \undefined \def \showISSN      #1{\unskip}     \fi
\ifx \showLCCN     \undefined \def \showLCCN      #1{\unskip}     \fi
\ifx \shownote     \undefined \def \shownote      #1{#1}          \fi
\ifx \showarticletitle \undefined \def \showarticletitle #1{#1}   \fi
\ifx \showURL      \undefined \def \showURL       {\relax}        \fi
\providecommand\bibfield[2]{#2}
\providecommand\bibinfo[2]{#2}
\providecommand\natexlab[1]{#1}
\providecommand\showeprint[2][]{arXiv:#2}

\bibitem[Akoury et~al\mbox{.}(2020)]%
        {akoury2020storium}
\bibfield{author}{\bibinfo{person}{Nader Akoury}, \bibinfo{person}{Shufan
  Wang}, \bibinfo{person}{Josh Whiting}, \bibinfo{person}{Stephen Hood},
  \bibinfo{person}{Nanyun Peng}, {and} \bibinfo{person}{Mohit Iyyer}.}
  \bibinfo{year}{2020}\natexlab{}.
\newblock \showarticletitle{Storium: A dataset and evaluation platform for
  machine-in-the-loop story generation}.
\newblock \bibinfo{journal}{\emph{arXiv preprint arXiv:2010.01717}}
  (\bibinfo{year}{2020}).
\newblock


\bibitem[Almasri et~al\mbox{.}(2019)]%
        {almasri2019intelligent}
\bibfield{author}{\bibinfo{person}{Abdelbaset Almasri}, \bibinfo{person}{Adel
  Ahmed}, \bibinfo{person}{Naser Almasri}, \bibinfo{person}{Yousef~S
  Abu~Sultan}, \bibinfo{person}{Ahmed~Y Mahmoud}, \bibinfo{person}{Ihab~S
  Zaqout}, \bibinfo{person}{Alaa~N Akkila}, {and} \bibinfo{person}{Samy~S
  Abu-Naser}.} \bibinfo{year}{2019}\natexlab{}.
\newblock \showarticletitle{Intelligent tutoring systems survey for the period
  2000-2018}.
\newblock  (\bibinfo{year}{2019}).
\newblock


\bibitem[Ballardini et~al\mbox{.}(2019)]%
        {ballardini2019ai}
\bibfield{author}{\bibinfo{person}{Rosa~Maria Ballardini}, \bibinfo{person}{Kan
  He}, {and} \bibinfo{person}{Teemu Roos}.} \bibinfo{year}{2019}\natexlab{}.
\newblock \showarticletitle{AI-generated content: authorship and inventorship
  in the age of artificial intelligence}.
\newblock \bibinfo{journal}{\emph{Online Distribution of Content in the EU}}
  (\bibinfo{year}{2019}), \bibinfo{pages}{117--135}.
\newblock


\bibitem[Bhat et~al\mbox{.}(2023)]%
        {bhat2023interacting}
\bibfield{author}{\bibinfo{person}{Advait Bhat}, \bibinfo{person}{Saaket
  Agashe}, \bibinfo{person}{Parth Oberoi}, \bibinfo{person}{Niharika Mohile},
  \bibinfo{person}{Ravi Jangir}, {and} \bibinfo{person}{Anirudha Joshi}.}
  \bibinfo{year}{2023}\natexlab{}.
\newblock \showarticletitle{Interacting with Next-Phrase Suggestions: How
  Suggestion Systems Aid and Influence the Cognitive Processes of Writing}. In
  \bibinfo{booktitle}{\emph{Proceedings of the 28th International Conference on
  Intelligent User Interfaces}} (Sydney, NSW, Australia)
  \emph{(\bibinfo{series}{IUI '23})}. \bibinfo{publisher}{Association for
  Computing Machinery}, \bibinfo{address}{New York, NY, USA},
  \bibinfo{pages}{436–452}.
\newblock
\showISBNx{9798400701061}
\urldef\tempurl%
\url{https://doi.org/10.1145/3581641.3584060}
\showDOI{\tempurl}


\bibitem[Blair(2011)]%
        {blair2011groundwork}
\bibfield{author}{\bibinfo{person}{J~Anthony Blair}.}
  \bibinfo{year}{2011}\natexlab{}.
\newblock \bibinfo{booktitle}{\emph{Groundwork in the theory of argumentation:
  Selected papers of J. Anthony Blair}}. Vol.~\bibinfo{volume}{21}.
\newblock \bibinfo{publisher}{Springer Science \& Business Media}.
\newblock


\bibitem[Booth et~al\mbox{.}(2003)]%
        {booth2003craft}
\bibfield{author}{\bibinfo{person}{Wayne~C Booth}, \bibinfo{person}{Gregory~G
  Colomb}, {and} \bibinfo{person}{Joseph~M Williams}.}
  \bibinfo{year}{2003}\natexlab{}.
\newblock \bibinfo{booktitle}{\emph{The craft of research}}.
\newblock \bibinfo{publisher}{University of Chicago press}.
\newblock


\bibitem[Borji(2023)]%
        {borji2023categorical}
\bibfield{author}{\bibinfo{person}{Ali Borji}.}
  \bibinfo{year}{2023}\natexlab{}.
\newblock \showarticletitle{A categorical archive of chatgpt failures}.
\newblock \bibinfo{journal}{\emph{arXiv preprint arXiv:2302.03494}}
  (\bibinfo{year}{2023}).
\newblock


\bibitem[Braun and Clarke(2006)]%
        {braun2006using}
\bibfield{author}{\bibinfo{person}{Virginia Braun} {and}
  \bibinfo{person}{Victoria Clarke}.} \bibinfo{year}{2006}\natexlab{}.
\newblock \showarticletitle{Using thematic analysis in psychology}.
\newblock \bibinfo{journal}{\emph{Qualitative research in psychology}}
  \bibinfo{volume}{3}, \bibinfo{number}{2} (\bibinfo{year}{2006}),
  \bibinfo{pages}{77--101}.
\newblock


\bibitem[Broll et~al\mbox{.}(2017)]%
        {broll2017visual}
\bibfield{author}{\bibinfo{person}{Brian Broll}, \bibinfo{person}{Akos
  L{\'e}deczi}, \bibinfo{person}{Peter Volgyesi}, \bibinfo{person}{Janos
  Sallai}, \bibinfo{person}{Miklos Maroti}, \bibinfo{person}{Alexia Carrillo},
  \bibinfo{person}{Stephanie~L Weeden-Wright}, \bibinfo{person}{Chris Vanags},
  \bibinfo{person}{Joshua~D Swartz}, {and} \bibinfo{person}{Melvin Lu}.}
  \bibinfo{year}{2017}\natexlab{}.
\newblock \showarticletitle{A visual programming environment for learning
  distributed programming}. In \bibinfo{booktitle}{\emph{Proceedings of the
  2017 ACM SIGCSE technical symposium on computer science education}}.
  \bibinfo{pages}{81--86}.
\newblock


\bibitem[Brown et~al\mbox{.}(2020)]%
        {brown2020language}
\bibfield{author}{\bibinfo{person}{Tom Brown}, \bibinfo{person}{Benjamin Mann},
  \bibinfo{person}{Nick Ryder}, \bibinfo{person}{Melanie Subbiah},
  \bibinfo{person}{Jared~D Kaplan}, \bibinfo{person}{Prafulla Dhariwal},
  \bibinfo{person}{Arvind Neelakantan}, \bibinfo{person}{Pranav Shyam},
  \bibinfo{person}{Girish Sastry}, \bibinfo{person}{Amanda Askell},
  {et~al\mbox{.}}} \bibinfo{year}{2020}\natexlab{}.
\newblock \showarticletitle{Language models are few-shot learners}.
\newblock \bibinfo{journal}{\emph{Advances in neural information processing
  systems}}  \bibinfo{volume}{33} (\bibinfo{year}{2020}),
  \bibinfo{pages}{1877--1901}.
\newblock


\bibitem[Chakrabarty et~al\mbox{.}(2022)]%
        {chakrabarty2022help}
\bibfield{author}{\bibinfo{person}{Tuhin Chakrabarty}, \bibinfo{person}{Vishakh
  Padmakumar}, {and} \bibinfo{person}{He He}.} \bibinfo{year}{2022}\natexlab{}.
\newblock \showarticletitle{Help me write a poem: Instruction Tuning as a
  Vehicle for Collaborative Poetry Writing}.
\newblock \bibinfo{journal}{\emph{arXiv preprint arXiv:2210.13669}}
  (\bibinfo{year}{2022}).
\newblock


\bibitem[Chen et~al\mbox{.}(2023)]%
        {chen2023teaching}
\bibfield{author}{\bibinfo{person}{Xinyun Chen}, \bibinfo{person}{Maxwell Lin},
  \bibinfo{person}{Nathanael Sch{\"a}rli}, {and} \bibinfo{person}{Denny Zhou}.}
  \bibinfo{year}{2023}\natexlab{}.
\newblock \showarticletitle{Teaching large language models to self-debug}.
\newblock \bibinfo{journal}{\emph{arXiv preprint arXiv:2304.05128}}
  (\bibinfo{year}{2023}).
\newblock


\bibitem[Chung et~al\mbox{.}(2022)]%
        {chung2022talebrush}
\bibfield{author}{\bibinfo{person}{John Joon~Young Chung},
  \bibinfo{person}{Wooseok Kim}, \bibinfo{person}{Kang~Min Yoo},
  \bibinfo{person}{Hwaran Lee}, \bibinfo{person}{Eytan Adar}, {and}
  \bibinfo{person}{Minsuk Chang}.} \bibinfo{year}{2022}\natexlab{}.
\newblock \showarticletitle{TaleBrush: sketching stories with generative
  pretrained language models}. In \bibinfo{booktitle}{\emph{Proceedings of the
  2022 CHI Conference on Human Factors in Computing Systems}}.
  \bibinfo{pages}{1--19}.
\newblock


\bibitem[Coenen et~al\mbox{.}(2021)]%
        {coenen2021wordcraft}
\bibfield{author}{\bibinfo{person}{Andy Coenen}, \bibinfo{person}{Luke Davis},
  \bibinfo{person}{Daphne Ippolito}, \bibinfo{person}{Emily Reif}, {and}
  \bibinfo{person}{Ann Yuan}.} \bibinfo{year}{2021}\natexlab{}.
\newblock \showarticletitle{Wordcraft: a human-ai collaborative editor for
  story writing}.
\newblock \bibinfo{journal}{\emph{arXiv preprint arXiv:2107.07430}}
  (\bibinfo{year}{2021}).
\newblock


\bibitem[Dang et~al\mbox{.}(2022)]%
        {dang2022beyond}
\bibfield{author}{\bibinfo{person}{Hai Dang}, \bibinfo{person}{Karim
  Benharrak}, \bibinfo{person}{Florian Lehmann}, {and} \bibinfo{person}{Daniel
  Buschek}.} \bibinfo{year}{2022}\natexlab{}.
\newblock \showarticletitle{Beyond Text Generation: Supporting Writers with
  Continuous Automatic Text Summaries}. In
  \bibinfo{booktitle}{\emph{Proceedings of the 35th Annual ACM Symposium on
  User Interface Software and Technology}}. \bibinfo{pages}{1--13}.
\newblock


\bibitem[Eemeren et~al\mbox{.}(1987)]%
        {eemeren1987handbook}
\bibfield{author}{\bibinfo{person}{Frans H~van Eemeren}, \bibinfo{person}{Rob
  Grootendorst}, {and} \bibinfo{person}{Tjark Kruiger}.}
  \bibinfo{year}{1987}\natexlab{}.
\newblock \bibinfo{booktitle}{\emph{Handbook of argumentation theory: A
  critical survey of classical backgrounds and modern studies}}.
\newblock \bibinfo{publisher}{De Gruyter}.
\newblock


\bibitem[Ferretti and Graham(2019)]%
        {ferretti2019argumentative}
\bibfield{author}{\bibinfo{person}{Ralph~P Ferretti} {and}
  \bibinfo{person}{Steve Graham}.} \bibinfo{year}{2019}\natexlab{}.
\newblock \showarticletitle{Argumentative writing: Theory, assessment, and
  instruction}.
\newblock \bibinfo{journal}{\emph{Reading and Writing}}  \bibinfo{volume}{32}
  (\bibinfo{year}{2019}), \bibinfo{pages}{1345--1357}.
\newblock


\bibitem[Ferretti and Lewis(2018)]%
        {ferretti2018argumentative}
\bibfield{author}{\bibinfo{person}{Ralph~P Ferretti} {and}
  \bibinfo{person}{William~E Lewis}.} \bibinfo{year}{2018}\natexlab{}.
\newblock \showarticletitle{Argumentative writing}.
\newblock \bibinfo{journal}{\emph{Best practices in writing instruction}}
  \bibinfo{volume}{135} (\bibinfo{year}{2018}).
\newblock


\bibitem[Flower and Hayes(1981)]%
        {flower1981cognitive}
\bibfield{author}{\bibinfo{person}{Linda Flower} {and} \bibinfo{person}{John~R
  Hayes}.} \bibinfo{year}{1981}\natexlab{}.
\newblock \showarticletitle{A cognitive process theory of writing}.
\newblock \bibinfo{journal}{\emph{College composition and communication}}
  \bibinfo{volume}{32}, \bibinfo{number}{4} (\bibinfo{year}{1981}),
  \bibinfo{pages}{365--387}.
\newblock


\bibitem[Gage(2006)]%
        {gage_2006}
\bibfield{author}{\bibinfo{person}{John~T. Gage}.}
  \bibinfo{year}{2006}\natexlab{}.
\newblock \bibinfo{booktitle}{\emph{The shape of reason: Argumentative writing
  in college}}.
\newblock \bibinfo{publisher}{Pearson Education}.
\newblock


\bibitem[Gao et~al\mbox{.}(2023)]%
        {gao2023collabcoder}
\bibfield{author}{\bibinfo{person}{Jie Gao}, \bibinfo{person}{Yuchen Guo},
  \bibinfo{person}{Gionnieve Lim}, \bibinfo{person}{Tianqin Zhang},
  \bibinfo{person}{Zheng Zhang}, \bibinfo{person}{Toby Jia-Jun Li}, {and}
  \bibinfo{person}{Simon~Tangi Perrault}.} \bibinfo{year}{2023}\natexlab{}.
\newblock \bibinfo{title}{CollabCoder: A GPT-Powered Workflow for Collaborative
  Qualitative Analysis}.
\newblock
\newblock
\showeprint[arxiv]{2304.07366}~[cs.HC]


\bibitem[Gero et~al\mbox{.}(2022)]%
        {gero2022sparks}
\bibfield{author}{\bibinfo{person}{Katy~Ilonka Gero}, \bibinfo{person}{Vivian
  Liu}, {and} \bibinfo{person}{Lydia Chilton}.}
  \bibinfo{year}{2022}\natexlab{}.
\newblock \showarticletitle{Sparks: Inspiration for science writing using
  language models}. In \bibinfo{booktitle}{\emph{Designing Interactive Systems
  Conference}}. \bibinfo{pages}{1002--1019}.
\newblock


\bibitem[Gollins and Gentner(2016)]%
        {gollins2016framework}
\bibfield{author}{\bibinfo{person}{Allan Gollins} {and} \bibinfo{person}{Dedre
  Gentner}.} \bibinfo{year}{2016}\natexlab{}.
\newblock \showarticletitle{A framework for a cognitive theory of writing}.
\newblock In \bibinfo{booktitle}{\emph{Cognitive processes in writing}}.
  \bibinfo{publisher}{Routledge}, \bibinfo{pages}{51--72}.
\newblock


\bibitem[Grabe and Kaplan(2014)]%
        {grabe2014theory}
\bibfield{author}{\bibinfo{person}{William Grabe} {and}
  \bibinfo{person}{Robert~B Kaplan}.} \bibinfo{year}{2014}\natexlab{}.
\newblock \bibinfo{booktitle}{\emph{Theory and practice of writing: An applied
  linguistic perspective}}.
\newblock \bibinfo{publisher}{Routledge}.
\newblock


\bibitem[Hayhoe et~al\mbox{.}(2003)]%
        {hayhoe2003visual}
\bibfield{author}{\bibinfo{person}{Mary~M Hayhoe}, \bibinfo{person}{Anurag
  Shrivastava}, \bibinfo{person}{Ryan Mruczek}, {and} \bibinfo{person}{Jeff~B
  Pelz}.} \bibinfo{year}{2003}\natexlab{}.
\newblock \showarticletitle{Visual memory and motor planning in a natural
  task}.
\newblock \bibinfo{journal}{\emph{Journal of vision}} \bibinfo{volume}{3},
  \bibinfo{number}{1} (\bibinfo{year}{2003}), \bibinfo{pages}{6--6}.
\newblock


\bibitem[Hitchcock and Verheij(2006)]%
        {hitchcock2006arguing}
\bibfield{author}{\bibinfo{person}{David Hitchcock} {and} \bibinfo{person}{Bart
  Verheij}.} \bibinfo{year}{2006}\natexlab{}.
\newblock \bibinfo{booktitle}{\emph{Arguing on the Toulmin model}}.
  Vol.~\bibinfo{volume}{10}.
\newblock \bibinfo{publisher}{Springer}.
\newblock


\bibitem[Hutchins et~al\mbox{.}(1985)]%
        {hutchins1985direct}
\bibfield{author}{\bibinfo{person}{Edwin~L Hutchins}, \bibinfo{person}{James~D
  Hollan}, {and} \bibinfo{person}{Donald~A Norman}.}
  \bibinfo{year}{1985}\natexlab{}.
\newblock \showarticletitle{Direct manipulation interfaces}.
\newblock \bibinfo{journal}{\emph{Human--computer interaction}}
  \bibinfo{volume}{1}, \bibinfo{number}{4} (\bibinfo{year}{1985}),
  \bibinfo{pages}{311--338}.
\newblock


\bibitem[Ippolito et~al\mbox{.}(2022)]%
        {ippolito2022creative}
\bibfield{author}{\bibinfo{person}{Daphne Ippolito}, \bibinfo{person}{Ann
  Yuan}, \bibinfo{person}{Andy Coenen}, {and} \bibinfo{person}{Sehmon Burnam}.}
  \bibinfo{year}{2022}\natexlab{}.
\newblock \showarticletitle{Creative Writing with an AI-Powered Writing
  Assistant: Perspectives from Professional Writers}.
\newblock \bibinfo{journal}{\emph{arXiv preprint arXiv:2211.05030}}
  (\bibinfo{year}{2022}).
\newblock


\bibitem[Johnson(2012)]%
        {johnson2012manifest}
\bibfield{author}{\bibinfo{person}{Ralph~H Johnson}.}
  \bibinfo{year}{2012}\natexlab{}.
\newblock \bibinfo{booktitle}{\emph{Manifest rationality: A pragmatic theory of
  argument}}.
\newblock \bibinfo{publisher}{Routledge}.
\newblock


\bibitem[Johnson-Laird(1999)]%
        {johnson1999deductive}
\bibfield{author}{\bibinfo{person}{Philip~N Johnson-Laird}.}
  \bibinfo{year}{1999}\natexlab{}.
\newblock \showarticletitle{Deductive reasoning}.
\newblock \bibinfo{journal}{\emph{Annual review of psychology}}
  \bibinfo{volume}{50}, \bibinfo{number}{1} (\bibinfo{year}{1999}),
  \bibinfo{pages}{109--135}.
\newblock


\bibitem[Kaddour et~al\mbox{.}(2023)]%
        {kaddour2023challenges}
\bibfield{author}{\bibinfo{person}{Jean Kaddour}, \bibinfo{person}{Joshua
  Harris}, \bibinfo{person}{Maximilian Mozes}, \bibinfo{person}{Herbie
  Bradley}, \bibinfo{person}{Roberta Raileanu}, {and} \bibinfo{person}{Robert
  McHardy}.} \bibinfo{year}{2023}\natexlab{}.
\newblock \showarticletitle{Challenges and Applications of Large Language
  Models}.
\newblock \bibinfo{journal}{\emph{arXiv preprint arXiv:2307.10169}}
  (\bibinfo{year}{2023}).
\newblock


\bibitem[Kahneman(2011)]%
        {kahneman2011thinking}
\bibfield{author}{\bibinfo{person}{Daniel Kahneman}.}
  \bibinfo{year}{2011}\natexlab{}.
\newblock \bibinfo{booktitle}{\emph{Thinking, fast and slow}}.
\newblock \bibinfo{publisher}{macmillan}.
\newblock


\bibitem[Kleemola et~al\mbox{.}(2022)]%
        {Kleemola2022TheCO}
\bibfield{author}{\bibinfo{person}{Katri Kleemola}, \bibinfo{person}{Heidi
  Hyytinen}, {and} \bibinfo{person}{Auli Toom}.}
  \bibinfo{year}{2022}\natexlab{}.
\newblock \showarticletitle{The Challenge of Position-Taking in Novice Higher
  Education Students’ Argumentative Writing}. In
  \bibinfo{booktitle}{\emph{Frontiers in Education}}.
\newblock


\bibitem[Kozbelt et~al\mbox{.}(2010)]%
        {kozbelt2010theories}
\bibfield{author}{\bibinfo{person}{Aaron Kozbelt}, \bibinfo{person}{Ronald~A
  Beghetto}, {and} \bibinfo{person}{Mark~A Runco}.}
  \bibinfo{year}{2010}\natexlab{}.
\newblock \showarticletitle{Theories of creativity.}
\newblock  (\bibinfo{year}{2010}).
\newblock


\bibitem[Kreminski and Martens(2022)]%
        {kreminski2022unmet}
\bibfield{author}{\bibinfo{person}{Max Kreminski} {and} \bibinfo{person}{Chris
  Martens}.} \bibinfo{year}{2022}\natexlab{}.
\newblock \showarticletitle{Unmet creativity support needs in computationally
  supported creative writing}. In \bibinfo{booktitle}{\emph{Proceedings of the
  First Workshop on Intelligent and Interactive Writing Assistants (In2Writing
  2022)}}. \bibinfo{pages}{74--82}.
\newblock


\bibitem[Landis and Koch(1977)]%
        {landis1977measurement}
\bibfield{author}{\bibinfo{person}{J~Richard Landis} {and}
  \bibinfo{person}{Gary~G Koch}.} \bibinfo{year}{1977}\natexlab{}.
\newblock \showarticletitle{The measurement of observer agreement for
  categorical data}.
\newblock \bibinfo{journal}{\emph{biometrics}} (\bibinfo{year}{1977}),
  \bibinfo{pages}{159--174}.
\newblock


\bibitem[Lawrence and Reed(2020)]%
        {lawrence2020argument}
\bibfield{author}{\bibinfo{person}{John Lawrence} {and} \bibinfo{person}{Chris
  Reed}.} \bibinfo{year}{2020}\natexlab{}.
\newblock \showarticletitle{Argument mining: A survey}.
\newblock \bibinfo{journal}{\emph{Computational Linguistics}}
  \bibinfo{volume}{45}, \bibinfo{number}{4} (\bibinfo{year}{2020}),
  \bibinfo{pages}{765--818}.
\newblock


\bibitem[Lazar et~al\mbox{.}(2017)]%
        {lazar2017research}
\bibfield{author}{\bibinfo{person}{Jonathan Lazar},
  \bibinfo{person}{Jinjuan~Heidi Feng}, {and} \bibinfo{person}{Harry
  Hochheiser}.} \bibinfo{year}{2017}\natexlab{}.
\newblock \bibinfo{booktitle}{\emph{Research methods in human-computer
  interaction}}.
\newblock \bibinfo{publisher}{Morgan Kaufmann}.
\newblock


\bibitem[Lee et~al\mbox{.}(2022b)]%
        {lee2022coauthor}
\bibfield{author}{\bibinfo{person}{Mina Lee}, \bibinfo{person}{Percy Liang},
  {and} \bibinfo{person}{Qian Yang}.} \bibinfo{year}{2022}\natexlab{b}.
\newblock \showarticletitle{Coauthor: Designing a human-ai collaborative
  writing dataset for exploring language model capabilities}. In
  \bibinfo{booktitle}{\emph{Proceedings of the 2022 CHI Conference on Human
  Factors in Computing Systems}}. \bibinfo{pages}{1--19}.
\newblock


\bibitem[Lee et~al\mbox{.}(2022a)]%
        {lee2022interactive}
\bibfield{author}{\bibinfo{person}{Yoonjoo Lee}, \bibinfo{person}{Tae~Soo Kim},
  \bibinfo{person}{Minsuk Chang}, {and} \bibinfo{person}{Juho Kim}.}
  \bibinfo{year}{2022}\natexlab{a}.
\newblock \showarticletitle{Interactive Children’s Story Rewriting Through
  Parent-Children Interaction}. In \bibinfo{booktitle}{\emph{Proceedings of the
  First Workshop on Intelligent and Interactive Writing Assistants (In2Writing
  2022)}}. \bibinfo{pages}{62--71}.
\newblock


\bibitem[Li et~al\mbox{.}(2020)]%
        {li_sovite:_2020}
\bibfield{author}{\bibinfo{person}{Toby Jia-Jun Li}, \bibinfo{person}{Jingya
  Chen}, \bibinfo{person}{Haijun Xia}, \bibinfo{person}{Tom~M. Mitchell}, {and}
  \bibinfo{person}{Brad~A. Myers}.} \bibinfo{year}{2020}\natexlab{}.
\newblock \showarticletitle{{Multi-Modal} {Repairs} of {Conversational}
  {Breakdowns} in {Task-Oriented} {Dialogs}}. In
  \bibinfo{booktitle}{\emph{Proceedings of the 33rd {Annual} {ACM} {Symposium}
  on {User} {Interface} {Software} and {Technology}}}
  \emph{(\bibinfo{series}{{UIST} 2020})}. \bibinfo{publisher}{ACM}.
\newblock
\urldef\tempurl%
\url{https://doi.org/10.1145/3379337.3415820}
\showDOI{\tempurl}


\bibitem[Li et~al\mbox{.}(2019)]%
        {li_pumice:_2019}
\bibfield{author}{\bibinfo{person}{Toby Jia-Jun Li}, \bibinfo{person}{Marissa
  Radensky}, \bibinfo{person}{Justin Jia}, \bibinfo{person}{Kirielle
  Singarajah}, \bibinfo{person}{Tom~M. Mitchell}, {and}
  \bibinfo{person}{Brad~A. Myers}.} \bibinfo{year}{2019}\natexlab{}.
\newblock \showarticletitle{{PUMICE}: {A} {Multi}-{Modal} {Agent} that {Learns}
  {Concepts} and {Conditionals} from {Natural} {Language} and
  {Demonstrations}}. In \bibinfo{booktitle}{\emph{Proceedings of the 32nd
  {Annual} {ACM} {Symposium} on {User} {Interface} {Software} and
  {Technology}}} \emph{(\bibinfo{series}{{UIST} 2019})}.
  \bibinfo{publisher}{ACM}.
\newblock
\urldef\tempurl%
\url{https://doi.org/10.1145/3332165.3347899}
\showDOI{\tempurl}


\bibitem[Lindgren and Sullivan(2006)]%
        {lindgren2006writing}
\bibfield{author}{\bibinfo{person}{Eva Lindgren} {and} \bibinfo{person}{Kirk~PH
  Sullivan}.} \bibinfo{year}{2006}\natexlab{}.
\newblock \showarticletitle{Writing and the analysis of revision: An overview}.
\newblock In \bibinfo{booktitle}{\emph{Computer key-stroke logging and
  writing}}. \bibinfo{publisher}{Brill}, \bibinfo{pages}{31--44}.
\newblock


\bibitem[Liu et~al\mbox{.}(2023)]%
        {liu2023pre}
\bibfield{author}{\bibinfo{person}{Pengfei Liu}, \bibinfo{person}{Weizhe Yuan},
  \bibinfo{person}{Jinlan Fu}, \bibinfo{person}{Zhengbao Jiang},
  \bibinfo{person}{Hiroaki Hayashi}, {and} \bibinfo{person}{Graham Neubig}.}
  \bibinfo{year}{2023}\natexlab{}.
\newblock \showarticletitle{Pre-train, prompt, and predict: A systematic survey
  of prompting methods in natural language processing}.
\newblock \bibinfo{journal}{\emph{Comput. Surveys}} \bibinfo{volume}{55},
  \bibinfo{number}{9} (\bibinfo{year}{2023}), \bibinfo{pages}{1--35}.
\newblock


\bibitem[Lovejoy(2011)]%
        {lovejoy2011great}
\bibfield{author}{\bibinfo{person}{Arthur~O Lovejoy}.}
  \bibinfo{year}{2011}\natexlab{}.
\newblock \bibinfo{booktitle}{\emph{The great chain of being: A study of the
  history of an idea}}.
\newblock \bibinfo{publisher}{Transaction Publishers}.
\newblock


\bibitem[Ltd.(2023)]%
        {ref-n-write_2023}
\bibfield{author}{\bibinfo{person}{Astute Digital~Solutions Ltd.}}
  \bibinfo{year}{2023}\natexlab{}.
\newblock \bibinfo{title}{Academic phrasebook - A guide for writing research
  papers}.
\newblock
\newblock
\urldef\tempurl%
\url{https://www.ref-n-write.com/academic-phrases-handbook/}
\showURL{%
\tempurl}


\bibitem[Lu et~al\mbox{.}(2022)]%
        {lu_bridging_2022}
\bibfield{author}{\bibinfo{person}{Yuwen Lu}, \bibinfo{person}{Chengzhi Zhang},
  \bibinfo{person}{Iris Zhang}, {and} \bibinfo{person}{Toby Jia-Jun Li}.}
  \bibinfo{year}{2022}\natexlab{}.
\newblock \showarticletitle{Bridging the Gap Between UX Practitioners’ Work
  Practices and AI-Enabled Design Support Tools}. In
  \bibinfo{booktitle}{\emph{Extended Abstracts of the 2022 CHI Conference on
  Human Factors in Computing Systems}} (New Orleans, LA, USA)
  \emph{(\bibinfo{series}{CHI EA '22})}. \bibinfo{publisher}{Association for
  Computing Machinery}, \bibinfo{address}{New York, NY, USA}, Article
  \bibinfo{articleno}{268}, \bibinfo{numpages}{7}~pages.
\newblock
\showISBNx{9781450391566}
\urldef\tempurl%
\url{https://doi.org/10.1145/3491101.3519809}
\showDOI{\tempurl}


\bibitem[Lunsford and Ruszkiewicz(2016)]%
        {lunsford2016everything}
\bibfield{author}{\bibinfo{person}{Andrea~A Lunsford} {and}
  \bibinfo{person}{John~J Ruszkiewicz}.} \bibinfo{year}{2016}\natexlab{}.
\newblock \bibinfo{booktitle}{\emph{Everything's an Argument}}.
\newblock \bibinfo{publisher}{Bedford/St. Martin's}.
\newblock


\bibitem[Lunzer and Hornb\ae{}k(2008)]%
        {lunzer_2008_subjective}
\bibfield{author}{\bibinfo{person}{Aran Lunzer} {and} \bibinfo{person}{Kasper
  Hornb\ae{}k}.} \bibinfo{year}{2008}\natexlab{}.
\newblock \showarticletitle{Subjunctive Interfaces: Extending Applications to
  Support Parallel Setup, Viewing and Control of Alternative Scenarios}.
\newblock \bibinfo{journal}{\emph{ACM Trans. Comput.-Hum. Interact.}}
  \bibinfo{volume}{14}, \bibinfo{number}{4}, Article \bibinfo{articleno}{17}
  (\bibinfo{date}{jan} \bibinfo{year}{2008}), \bibinfo{numpages}{44}~pages.
\newblock
\showISSN{1073-0516}
\urldef\tempurl%
\url{https://doi.org/10.1145/1314683.1314685}
\showDOI{\tempurl}


\bibitem[McAlister et~al\mbox{.}(2017)]%
        {mcalister2017qualitative}
\bibfield{author}{\bibinfo{person}{Anne~Marguerite McAlister},
  \bibinfo{person}{Dennis~M Lee}, \bibinfo{person}{Katherine~M Ehlert},
  \bibinfo{person}{Rachel~Louis Kajfez}, \bibinfo{person}{Courtney~June Faber},
  {and} \bibinfo{person}{Marian~S Kennedy}.} \bibinfo{year}{2017}\natexlab{}.
\newblock \showarticletitle{Qualitative coding: An approach to assess
  inter-rater reliability}. In \bibinfo{booktitle}{\emph{2017 ASEE annual
  conference \& exposition}}.
\newblock


\bibitem[M{\"u}ller-Wienbergen et~al\mbox{.}(2011)]%
        {muller2011leaving}
\bibfield{author}{\bibinfo{person}{Felix M{\"u}ller-Wienbergen},
  \bibinfo{person}{Oliver M{\"u}ller}, \bibinfo{person}{Stefan Seidel}, {and}
  \bibinfo{person}{J{\"o}rg Becker}.} \bibinfo{year}{2011}\natexlab{}.
\newblock \showarticletitle{Leaving the beaten tracks in creative work--A
  design theory for systems that support convergent and divergent thinking}.
\newblock \bibinfo{journal}{\emph{Journal of the Association for Information
  Systems}} \bibinfo{volume}{12}, \bibinfo{number}{11} (\bibinfo{year}{2011}),
  \bibinfo{pages}{2}.
\newblock


\bibitem[Myers(1986)]%
        {myers1986visual}
\bibfield{author}{\bibinfo{person}{Brad~A Myers}.}
  \bibinfo{year}{1986}\natexlab{}.
\newblock \showarticletitle{Visual programming, programming by example, and
  program visualization: a taxonomy}.
\newblock \bibinfo{journal}{\emph{ACM sigchi bulletin}} \bibinfo{volume}{17},
  \bibinfo{number}{4} (\bibinfo{year}{1986}), \bibinfo{pages}{59--66}.
\newblock


\bibitem[Nelson(1998)]%
        {nelson1998theorizing}
\bibfield{author}{\bibinfo{person}{Emmanuel~Sampath Nelson}.}
  \bibinfo{year}{1998}\natexlab{}.
\newblock \bibinfo{booktitle}{\emph{Theorizing composition: A critical
  sourcebook of theory and scholarship in contemporary composition studies}}.
\newblock \bibinfo{publisher}{Greenwood Publishing Group}.
\newblock


\bibitem[Pedemonte(2007)]%
        {pedemonte2007can}
\bibfield{author}{\bibinfo{person}{Bettina Pedemonte}.}
  \bibinfo{year}{2007}\natexlab{}.
\newblock \showarticletitle{How can the relationship between argumentation and
  proof be analysed?}
\newblock \bibinfo{journal}{\emph{Educational studies in mathematics}}
  \bibinfo{volume}{66}, \bibinfo{number}{1} (\bibinfo{year}{2007}),
  \bibinfo{pages}{23--41}.
\newblock


\bibitem[Rapp(2002)]%
        {rapp2002aristotle}
\bibfield{author}{\bibinfo{person}{Christof Rapp}.}
  \bibinfo{year}{2002}\natexlab{}.
\newblock \showarticletitle{Aristotle’s rhetoric}.
\newblock  (\bibinfo{year}{2002}).
\newblock


\bibitem[Raza et~al\mbox{.}(2022)]%
        {raza2022dbias}
\bibfield{author}{\bibinfo{person}{Shaina Raza}, \bibinfo{person}{Deepak~John
  Reji}, {and} \bibinfo{person}{Chen Ding}.} \bibinfo{year}{2022}\natexlab{}.
\newblock \showarticletitle{Dbias: detecting biases and ensuring fairness in
  news articles}.
\newblock \bibinfo{journal}{\emph{International Journal of Data Science and
  Analytics}} (\bibinfo{year}{2022}), \bibinfo{pages}{1--21}.
\newblock


\bibitem[Reynolds and McDonell(2021)]%
        {reynolds2021prompt}
\bibfield{author}{\bibinfo{person}{Laria Reynolds} {and} \bibinfo{person}{Kyle
  McDonell}.} \bibinfo{year}{2021}\natexlab{}.
\newblock \showarticletitle{Prompt programming for large language models:
  Beyond the few-shot paradigm}. In \bibinfo{booktitle}{\emph{Extended
  Abstracts of the 2021 CHI Conference on Human Factors in Computing Systems}}.
  \bibinfo{pages}{1--7}.
\newblock


\bibitem[Ruiz-Dolz et~al\mbox{.}(2022)]%
        {ruiz2022qualitative}
\bibfield{author}{\bibinfo{person}{Ramon Ruiz-Dolz}, \bibinfo{person}{Joaquin
  Taverner}, \bibinfo{person}{Stella Heras}, \bibinfo{person}{Ana
  Garcia-Fornes}, {and} \bibinfo{person}{Vicente Botti}.}
  \bibinfo{year}{2022}\natexlab{}.
\newblock \showarticletitle{A Qualitative Analysis of the Persuasive Properties
  of Argumentation Schemes}. In \bibinfo{booktitle}{\emph{Proceedings of the
  30th ACM Conference on User Modeling, Adaptation and Personalization}}.
  \bibinfo{pages}{1--11}.
\newblock


\bibitem[Runco and Jaeger(2012)]%
        {runco2012standard}
\bibfield{author}{\bibinfo{person}{Mark~A Runco} {and}
  \bibinfo{person}{Garrett~J Jaeger}.} \bibinfo{year}{2012}\natexlab{}.
\newblock \showarticletitle{The standard definition of creativity}.
\newblock \bibinfo{journal}{\emph{Creativity research journal}}
  \bibinfo{volume}{24}, \bibinfo{number}{1} (\bibinfo{year}{2012}),
  \bibinfo{pages}{92--96}.
\newblock


\bibitem[Ryan and Morgan(2007)]%
        {ryan2007modern}
\bibfield{author}{\bibinfo{person}{Thomas~P Ryan} {and} \bibinfo{person}{JP
  Morgan}.} \bibinfo{year}{2007}\natexlab{}.
\newblock \showarticletitle{Modern experimental design}.
\newblock \bibinfo{journal}{\emph{Journal of Statistical Theory and Practice}}
  \bibinfo{volume}{1}, \bibinfo{number}{3-4} (\bibinfo{year}{2007}),
  \bibinfo{pages}{501--506}.
\newblock


\bibitem[Saparov and He(2022)]%
        {saparov2022language}
\bibfield{author}{\bibinfo{person}{Abulhair Saparov} {and} \bibinfo{person}{He
  He}.} \bibinfo{year}{2022}\natexlab{}.
\newblock \showarticletitle{Language models are greedy reasoners: A systematic
  formal analysis of chain-of-thought}.
\newblock \bibinfo{journal}{\emph{arXiv preprint arXiv:2210.01240}}
  (\bibinfo{year}{2022}).
\newblock


\bibitem[Schwarz et~al\mbox{.}(2017)]%
        {schwarz2017dialogue}
\bibfield{author}{\bibinfo{person}{Baruch~B Schwarz}, \bibinfo{person}{Lauren~B
  Resnick}, {and} \bibinfo{person}{Michael~J Baker}.}
  \bibinfo{year}{2017}\natexlab{}.
\newblock \bibinfo{booktitle}{\emph{Dialogue, argumentation and education:
  History, theory and practice}}.
\newblock \bibinfo{publisher}{Cambridge University Press}.
\newblock


\bibitem[Shinn et~al\mbox{.}(2023)]%
        {shinn2023reflexion}
\bibfield{author}{\bibinfo{person}{Noah Shinn}, \bibinfo{person}{Beck Labash},
  {and} \bibinfo{person}{Ashwin Gopinath}.} \bibinfo{year}{2023}\natexlab{}.
\newblock \showarticletitle{Reflexion: an autonomous agent with dynamic memory
  and self-reflection}.
\newblock \bibinfo{journal}{\emph{arXiv preprint arXiv:2303.11366}}
  (\bibinfo{year}{2023}).
\newblock


\bibitem[Shneiderman and Maes(1997)]%
        {shneiderman_1997_direct}
\bibfield{author}{\bibinfo{person}{Ben Shneiderman} {and}
  \bibinfo{person}{Pattie Maes}.} \bibinfo{year}{1997}\natexlab{}.
\newblock \showarticletitle{Direct Manipulation vs. Interface Agents}.
\newblock \bibinfo{journal}{\emph{Interactions}} \bibinfo{volume}{4},
  \bibinfo{number}{6} (\bibinfo{date}{nov} \bibinfo{year}{1997}),
  \bibinfo{pages}{42–61}.
\newblock
\showISSN{1072-5520}
\urldef\tempurl%
\url{https://doi.org/10.1145/267505.267514}
\showDOI{\tempurl}


\bibitem[Singh et~al\mbox{.}(2022)]%
        {singh2022hide}
\bibfield{author}{\bibinfo{person}{Nikhil Singh}, \bibinfo{person}{Guillermo
  Bernal}, \bibinfo{person}{Daria Savchenko}, {and} \bibinfo{person}{Elena~L
  Glassman}.} \bibinfo{year}{2022}\natexlab{}.
\newblock \showarticletitle{Where to hide a stolen elephant: Leaps in creative
  writing with multimodal machine intelligence}.
\newblock \bibinfo{journal}{\emph{ACM Transactions on Computer-Human
  Interaction}} (\bibinfo{year}{2022}).
\newblock


\bibitem[Snyder(2019)]%
        {snyder2019literature}
\bibfield{author}{\bibinfo{person}{Hannah Snyder}.}
  \bibinfo{year}{2019}\natexlab{}.
\newblock \showarticletitle{Literature review as a research methodology: An
  overview and guidelines}.
\newblock \bibinfo{journal}{\emph{Journal of business research}}
  \bibinfo{volume}{104} (\bibinfo{year}{2019}), \bibinfo{pages}{333--339}.
\newblock


\bibitem[Suh et~al\mbox{.}(2022)]%
        {suh2022codetoon}
\bibfield{author}{\bibinfo{person}{Sangho Suh}, \bibinfo{person}{Jian Zhao},
  {and} \bibinfo{person}{Edith Law}.} \bibinfo{year}{2022}\natexlab{}.
\newblock \showarticletitle{Codetoon: Story ideation, auto comic generation,
  and structure mapping for code-driven storytelling}. In
  \bibinfo{booktitle}{\emph{Proceedings of the 35th Annual ACM Symposium on
  User Interface Software and Technology}}. \bibinfo{pages}{1--16}.
\newblock


\bibitem[Suzgun et~al\mbox{.}(2022)]%
        {suzgun2022challenging}
\bibfield{author}{\bibinfo{person}{Mirac Suzgun}, \bibinfo{person}{Nathan
  Scales}, \bibinfo{person}{Nathanael Sch{\"a}rli}, \bibinfo{person}{Sebastian
  Gehrmann}, \bibinfo{person}{Yi Tay}, \bibinfo{person}{Hyung~Won Chung},
  \bibinfo{person}{Aakanksha Chowdhery}, \bibinfo{person}{Quoc~V Le},
  \bibinfo{person}{Ed~H Chi}, \bibinfo{person}{Denny Zhou}, {et~al\mbox{.}}}
  \bibinfo{year}{2022}\natexlab{}.
\newblock \showarticletitle{Challenging BIG-Bench tasks and whether
  chain-of-thought can solve them}.
\newblock \bibinfo{journal}{\emph{arXiv preprint arXiv:2210.09261}}
  (\bibinfo{year}{2022}).
\newblock


\bibitem[Swenson(1993)]%
        {swenson1993visual}
\bibfield{author}{\bibinfo{person}{Keith~D Swenson}.}
  \bibinfo{year}{1993}\natexlab{}.
\newblock \showarticletitle{A visual language to describe collaborative work}.
  In \bibinfo{booktitle}{\emph{Proceedings 1993 IEEE Symposium on Visual
  Languages}}. IEEE, \bibinfo{pages}{298--303}.
\newblock


\bibitem[Van~Eemeren et~al\mbox{.}(2013)]%
        {van2013fundamentals}
\bibfield{author}{\bibinfo{person}{Frans~H Van~Eemeren}, \bibinfo{person}{Rob
  Grootendorst}, \bibinfo{person}{Ralph~H Johnson}, \bibinfo{person}{Christian
  Plantin}, {and} \bibinfo{person}{Charles~A Willard}.}
  \bibinfo{year}{2013}\natexlab{}.
\newblock \bibinfo{booktitle}{\emph{Fundamentals of argumentation theory: A
  handbook of historical backgrounds and contemporary developments}}.
\newblock \bibinfo{publisher}{Routledge}.
\newblock


\bibitem[Wachsmuth et~al\mbox{.}(2017)]%
        {wachsmuth2017computational}
\bibfield{author}{\bibinfo{person}{Henning Wachsmuth}, \bibinfo{person}{Nona
  Naderi}, \bibinfo{person}{Yufang Hou}, \bibinfo{person}{Yonatan Bilu},
  \bibinfo{person}{Vinodkumar Prabhakaran}, \bibinfo{person}{Tim~Alberdingk
  Thijm}, \bibinfo{person}{Graeme Hirst}, {and} \bibinfo{person}{Benno Stein}.}
  \bibinfo{year}{2017}\natexlab{}.
\newblock \showarticletitle{Computational argumentation quality assessment in
  natural language}. In \bibinfo{booktitle}{\emph{Proceedings of the 15th
  Conference of the European Chapter of the Association for Computational
  Linguistics: Volume 1, Long Papers}}. \bibinfo{pages}{176--187}.
\newblock


\bibitem[Wambsganss et~al\mbox{.}(2021)]%
        {wambsganss2021arguetutor}
\bibfield{author}{\bibinfo{person}{Thiemo Wambsganss}, \bibinfo{person}{Tobias
  Kueng}, \bibinfo{person}{Matthias Soellner}, {and} \bibinfo{person}{Jan~Marco
  Leimeister}.} \bibinfo{year}{2021}\natexlab{}.
\newblock \showarticletitle{ArgueTutor: An adaptive dialog-based learning
  system for argumentation skills}. In \bibinfo{booktitle}{\emph{Proceedings of
  the 2021 CHI conference on human factors in computing systems}}.
  \bibinfo{pages}{1--13}.
\newblock


\bibitem[Wambsganss et~al\mbox{.}(2020)]%
        {wambsganss2020adaptive}
\bibfield{author}{\bibinfo{person}{Thiemo Wambsganss},
  \bibinfo{person}{Christina Niklaus}, \bibinfo{person}{Matthias Cetto},
  \bibinfo{person}{Matthias S{\"o}llner}, \bibinfo{person}{Siegfried
  Handschuh}, {and} \bibinfo{person}{Jan~Marco Leimeister}.}
  \bibinfo{year}{2020}\natexlab{}.
\newblock \showarticletitle{AL: An adaptive learning support system for
  argumentation skills}. In \bibinfo{booktitle}{\emph{Proceedings of the 2020
  CHI Conference on Human Factors in Computing Systems}}.
  \bibinfo{pages}{1--14}.
\newblock


\bibitem[Wang et~al\mbox{.}(2015)]%
        {wang2015docuviz}
\bibfield{author}{\bibinfo{person}{Dakuo Wang}, \bibinfo{person}{Judith~S
  Olson}, \bibinfo{person}{Jingwen Zhang}, \bibinfo{person}{Trung Nguyen},
  {and} \bibinfo{person}{Gary~M Olson}.} \bibinfo{year}{2015}\natexlab{}.
\newblock \showarticletitle{DocuViz: visualizing collaborative writing}. In
  \bibinfo{booktitle}{\emph{Proceedings of the 33rd Annual ACM conference on
  human factors in computing systems}}. \bibinfo{pages}{1865--1874}.
\newblock


\bibitem[Wang et~al\mbox{.}(2017)]%
        {wang2017users}
\bibfield{author}{\bibinfo{person}{Dakuo Wang}, \bibinfo{person}{Haodan Tan},
  {and} \bibinfo{person}{Tun Lu}.} \bibinfo{year}{2017}\natexlab{}.
\newblock \showarticletitle{Why users do not want to write together when they
  are writing together: Users' rationales for today's collaborative writing
  practices}.
\newblock \bibinfo{journal}{\emph{Proceedings of the ACM on Human-Computer
  Interaction}} \bibinfo{volume}{1}, \bibinfo{number}{CSCW}
  (\bibinfo{year}{2017}), \bibinfo{pages}{1--18}.
\newblock


\bibitem[Wei et~al\mbox{.}(2022)]%
        {wei2022chain}
\bibfield{author}{\bibinfo{person}{Jason Wei}, \bibinfo{person}{Xuezhi Wang},
  \bibinfo{person}{Dale Schuurmans}, \bibinfo{person}{Maarten Bosma},
  \bibinfo{person}{Ed Chi}, \bibinfo{person}{Quoc Le}, {and}
  \bibinfo{person}{Denny Zhou}.} \bibinfo{year}{2022}\natexlab{}.
\newblock \showarticletitle{Chain of thought prompting elicits reasoning in
  large language models}.
\newblock \bibinfo{journal}{\emph{arXiv preprint arXiv:2201.11903}}
  (\bibinfo{year}{2022}).
\newblock


\bibitem[Wentzel(2017)]%
        {wentzel2017guide}
\bibfield{author}{\bibinfo{person}{Arnold Wentzel}.}
  \bibinfo{year}{2017}\natexlab{}.
\newblock \bibinfo{booktitle}{\emph{A guide to argumentative research writing
  and thinking: Overcoming challenges}}.
\newblock \bibinfo{publisher}{Routledge}.
\newblock


\bibitem[Whitley(1997)]%
        {whitley1997visual}
\bibfield{author}{\bibinfo{person}{Kirsten~N. Whitley}.}
  \bibinfo{year}{1997}\natexlab{}.
\newblock \showarticletitle{Visual programming languages and the empirical
  evidence for and against}.
\newblock \bibinfo{journal}{\emph{Journal of Visual Languages \& Computing}}
  \bibinfo{volume}{8}, \bibinfo{number}{1} (\bibinfo{year}{1997}),
  \bibinfo{pages}{109--142}.
\newblock


\bibitem[Williams(2012)]%
        {williams_three_2012}
\bibfield{author}{\bibinfo{person}{Joseph~M. Williams}.}
  \bibinfo{year}{2012}\natexlab{}.
\newblock \bibinfo{booktitle}{\emph{Three {Modules} on {Clear} {Writing}
  {Style}: {An} {Introduction} to the {Craft} of {Argument}}}.
\newblock \bibinfo{publisher}{SOAR: Student Observation and Research}.
\newblock
\showISBNx{9780692254691}


\bibitem[Wong-Villacres et~al\mbox{.}(2015)]%
        {wong2015tabletop}
\bibfield{author}{\bibinfo{person}{Marisol Wong-Villacres},
  \bibinfo{person}{Margarita Ortiz}, \bibinfo{person}{Vanessa Echeverr{\'\i}a},
  {and} \bibinfo{person}{Katherine Chiluiza}.} \bibinfo{year}{2015}\natexlab{}.
\newblock \showarticletitle{A tabletop system to promote argumentation in
  computer science students}. In \bibinfo{booktitle}{\emph{Proceedings of the
  2015 International Conference on Interactive Tabletops \& Surfaces}}.
  \bibinfo{pages}{325--330}.
\newblock


\bibitem[Xia et~al\mbox{.}(2022)]%
        {xia2022persua}
\bibfield{author}{\bibinfo{person}{Meng Xia}, \bibinfo{person}{Qian Zhu},
  \bibinfo{person}{Xingbo Wang}, \bibinfo{person}{Fei Nie},
  \bibinfo{person}{Huamin Qu}, {and} \bibinfo{person}{Xiaojuan Ma}.}
  \bibinfo{year}{2022}\natexlab{}.
\newblock \showarticletitle{Persua: A visual interactive system to enhance the
  persuasiveness of arguments in online discussion}.
\newblock \bibinfo{journal}{\emph{Proceedings of the ACM on Human-Computer
  Interaction}} \bibinfo{volume}{6}, \bibinfo{number}{CSCW2}
  (\bibinfo{year}{2022}), \bibinfo{pages}{1--30}.
\newblock


\bibitem[Yang et~al\mbox{.}(2022b)]%
        {yang2022ai}
\bibfield{author}{\bibinfo{person}{Daijin Yang}, \bibinfo{person}{Yanpeng
  Zhou}, \bibinfo{person}{Zhiyuan Zhang}, \bibinfo{person}{Toby Jia-Jun Li},
  {and} \bibinfo{person}{Ray LC}.} \bibinfo{year}{2022}\natexlab{b}.
\newblock \showarticletitle{AI as an Active Writer: Interaction strategies with
  generated text in human-AI collaborative fiction writing}. In
  \bibinfo{booktitle}{\emph{Joint Proceedings of the ACM IUI Workshops}},
  Vol.~\bibinfo{volume}{10}.
\newblock


\bibitem[Yang et~al\mbox{.}(2022a)]%
        {yang2022re3}
\bibfield{author}{\bibinfo{person}{Kevin Yang}, \bibinfo{person}{Nanyun Peng},
  \bibinfo{person}{Yuandong Tian}, {and} \bibinfo{person}{Dan Klein}.}
  \bibinfo{year}{2022}\natexlab{a}.
\newblock \showarticletitle{Re3: Generating longer stories with recursive
  reprompting and revision}.
\newblock \bibinfo{journal}{\emph{arXiv preprint arXiv:2210.06774}}
  (\bibinfo{year}{2022}).
\newblock


\bibitem[Yao et~al\mbox{.}(2023)]%
        {yao2023tree}
\bibfield{author}{\bibinfo{person}{Shunyu Yao}, \bibinfo{person}{Dian Yu},
  \bibinfo{person}{Jeffrey Zhao}, \bibinfo{person}{Izhak Shafran},
  \bibinfo{person}{Thomas~L Griffiths}, \bibinfo{person}{Yuan Cao}, {and}
  \bibinfo{person}{Karthik Narasimhan}.} \bibinfo{year}{2023}\natexlab{}.
\newblock \showarticletitle{Tree of thoughts: Deliberate problem solving with
  large language models}.
\newblock \bibinfo{journal}{\emph{arXiv preprint arXiv:2305.10601}}
  (\bibinfo{year}{2023}).
\newblock


\bibitem[Yim et~al\mbox{.}(2017)]%
        {yim2017synchronous}
\bibfield{author}{\bibinfo{person}{Soobin Yim}, \bibinfo{person}{Dakuo Wang},
  \bibinfo{person}{Judith Olson}, \bibinfo{person}{Viet Vu}, {and}
  \bibinfo{person}{Mark Warschauer}.} \bibinfo{year}{2017}\natexlab{}.
\newblock \showarticletitle{Synchronous collaborative writing in the classroom:
  undergraduates' collaboration practices and their impact on writing style,
  quality, and quantity}. In \bibinfo{booktitle}{\emph{Proceedings of the 2017
  ACM Conference on Computer Supported Cooperative Work and Social Computing}}.
  \bibinfo{pages}{468--479}.
\newblock


\bibitem[Young et~al\mbox{.}(1995)]%
        {young1995cantata}
\bibfield{author}{\bibinfo{person}{Mark Young}, \bibinfo{person}{Danielle
  Argiro}, {and} \bibinfo{person}{Steven Kubica}.}
  \bibinfo{year}{1995}\natexlab{}.
\newblock \showarticletitle{Cantata: Visual programming environment for the
  Khoros system}.
\newblock \bibinfo{journal}{\emph{ACM SIGGRAPH Computer Graphics}}
  \bibinfo{volume}{29}, \bibinfo{number}{2} (\bibinfo{year}{1995}),
  \bibinfo{pages}{22--24}.
\newblock


\bibitem[Zhang et~al\mbox{.}(2022)]%
        {zhang2022storybuddy}
\bibfield{author}{\bibinfo{person}{Zheng Zhang}, \bibinfo{person}{Ying Xu},
  \bibinfo{person}{Yanhao Wang}, \bibinfo{person}{Bingsheng Yao},
  \bibinfo{person}{Daniel Ritchie}, \bibinfo{person}{Tongshuang Wu},
  \bibinfo{person}{Mo Yu}, \bibinfo{person}{Dakuo Wang}, {and}
  \bibinfo{person}{Toby Jia-Jun Li}.} \bibinfo{year}{2022}\natexlab{}.
\newblock \showarticletitle{Storybuddy: A human-ai collaborative chatbot for
  parent-child interactive storytelling with flexible parental involvement}. In
  \bibinfo{booktitle}{\emph{Proceedings of the 2022 CHI Conference on Human
  Factors in Computing Systems}}. \bibinfo{pages}{1--21}.
\newblock


\end{thebibliography}

\newpage
\appendix

\newcolumntype{T}{>{\centering\arraybackslash}m{0.1\linewidth}}
\newcolumntype{Y}{>{\centering\arraybackslash}m{0.5\linewidth}}
\newcolumntype{Z}{>{\centering\arraybackslash}m{0.3\linewidth}}

\begin{table*}[htbp]
\centering

\renewcommand{\arraystretch}{2.6}
\begin{tabularx}{\linewidth}{|T|Y|Z|}
\cline{1-3}
 & \textbf{Topic} & \textbf{Task description} \\
\cline{1-3}
Topic 1 & Governments should not fund any scientific research whose consequences are unclear & \multirow{3}{\linewidth}{Write a response in which you discuss the extent to which you agree or disagree with the statement and explain your reasoning for the position you take. In developing and supporting your position, you should consider ways in which the statement might or might not hold true and explain how these considerations shape your position} \\
\cline{1-2}
Topic 2 & People's attitudes are determined more by their immediate situation or surroundings than by society as a whole & \\
\cline{1-2}
Topic 3 & Educational institutions should actively encourage their students to choose fields of study that will prepare them for lucrative careers & \\
\cline{1-3}
\end{tabularx}
\label{table:topic}
\captionsetup{position=below}
\caption{Argumentative topics used in our user study, adopted from the sample topics of GRE issue writing}
\end{table*}

\newcolumntype{T}{>{\centering\arraybackslash}m{0.1\linewidth}}
\newcolumntype{Y}{>{\centering\arraybackslash}m{0.3\linewidth}}
\newcolumntype{Z}{>{\centering\arraybackslash}m{0.5\linewidth}}

\begin{table*}[]
\small

\centering

\renewcommand{\arraystretch}{3}
\begin{tabularx}{\linewidth}{|T|Y|Z|}
\cline{1-3}
  \textbf{Task} & \textbf{Prompt template} & \textbf{Few-shot examples} \\
\cline{1-3}

Key aspect generation & \begin{itemize} [leftmargin=*, itemindent=0pt, nosep]
    \vspace{\baselineskip}
    \item  \textbf{system role}: A helpful writing assistant that aims to help writers come up with high level aspects or topics that they can think of to support their argument
    \item \textbf{user input}: Please list key aspects that are worth discussing to support the argument: [selected argument]
\end{itemize} & 
Key aspects of the argument ``\textit{We should advocate for the expansion of engineering course offerings in colleges}'':
\begin{itemize}
    \item Growing demand for engineers
    \item Diverse career opportunities
    \item Promoting interdisciplinary learning
    \item Fostering innovation and creativity
    \item Enhancing STEM education
\end{itemize}
\\
\cline{1-3}

Discussion point generation & \begin{itemize} [leftmargin=*, itemindent=0pt, nosep]
    \vspace{\baselineskip}
    \item  \textbf{system role}: A helpful writing assistant that aims to come up with pertinent discussion points based
on a specified aspect to reinforce the given argument
    \item \textbf{user input}: Please list key discussion points worth including in the discussion to support argument [selected argument] from the aspect of [selected aspect]
\end{itemize} & 
Discussion points of the argument ``\textit{We should advocate for the expansion of engineering course offerings in colleges}'' from the perspective of ``\textit{Growing demand for engineers}'':
\begin{itemize}
    \item Recent increased job position number in Engineering
    \item National interest in key engineering area
    \item Varying engineering job types
    \item Labor scarcity in the engineering job market
\end{itemize}
\\
\cline{1-3}

Counter argument generation & \begin{itemize} [leftmargin=*, itemindent=0pt, nosep]
    \item  \textbf{system role}: A helpful writing assistant specializing in argumentative essay tutoring and generating counterarguments for the given statement
    \item \textbf{user input}: Please list potential counterarguments that can challenge
the argument [selected argument]
\end{itemize} & 
Counterarguments of the statement ``\textit{Some scholars and researchers claim that there are negative impacts of technology on a child’s developing mind. According to one research study, scholars claimed that “moderate evidence also suggests that early exposure to purely entertainment content, and media violence in particular, is negatively associated with cognitive skills and academic achievement}'':

\begin{itemize}
\item \textit{Evidence of Positive Impacts}: While some studies suggest negative impacts of technology on children's cognitive skills and academic achievement, there are also studies that demonstrate positive impacts, such as improved visual-spatial skills, problem-solving abilities, and creativity.

\item \textit{Importance of Parental Involvement}: The negative impacts of technology can be mitigated by parental involvement and guidance. Parents can set limits on screen time and select age-appropriate content, and monitor their children's technology use to ensure they are engaging in positive and educational activities.

\item \textit{Individual Differences}: Not all children are affected the same way by technology, and the impact on their cognitive skills and academic achievement may depend on individual factors, such as age, gender, socioeconomic status, and learning style.

\item \textit{Importance of Context}: The negative impacts of technology on cognitive skills and academic achievement may be dependent on the context in which it is used. For example, technology use in the classroom may have different effects than technology use at home.
\end{itemize}
\\

\cline{1-3}
\end{tabularx}
\label{table:topic}
\captionsetup{position=below}
\end{table*}

\begin{table*}[]
\small

\centering

\renewcommand{\arraystretch}{3}
\begin{tabularx}{\linewidth}{|T|Y|Z|}
\cline{1-3}

Logical fallacy generation & \begin{itemize} [leftmargin=*, itemindent=0pt, nosep]
    \item  \textbf{system role}: A helpful writing assistant focusing on argumentative essay tutoring and trying to suggest logical weaknesses in the given statement
    \item \textbf{user input}: Please list potential logical weaknesses in the argument [selected argument]
\end{itemize} & 

Logical fallacies in the statement ``\textit{The seriousness of a punishment should match the seriousness of the crime. Right now, the punishment for drunk driving may simply be a fine. But drunk driving is a very serious crime that can kill innocent people. So the death penalty should be the punishment for drunk driving}'':

\begin{itemize}
\item \textit{Slippery Slope Fallacy}: The argument assumes that the punishment for drunk driving should be escalated all the way to the death penalty, without considering other proportional punishments that could be implemented between a fine and the death penalty. This is a slippery slope fallacy.

\item \textit{False Analogy Fallacy}: The argument equates drunk driving, which is a serious crime that can result in innocent deaths, with other crimes that are punishable by the death penalty, such as murder. This is a false analogy fallacy, as drunk driving and murder are not equivalent in terms of their severity, intent, or harm caused.

\end{itemize}

Logical fallacies in the statement ``\textit{I know the exam is graded based on performance, but you should give me an A. My cat has been sick, my car broke down, and I’ve had a cold, so it was really hard for me to study!}'':

\begin{itemize}
\item \textit{Appeal to Pity Fallacy}: The argument attempts to persuade the grader by appealing to pity, by suggesting that the student's unfortunate circumstances should override their actual performance on the exam.

\item \textit{False Cause Fallacy}: The argument implies that the student's poor performance on the exam is caused by their external circumstances, such as a sick cat or a broken car, without providing any evidence to support this claim. This is a false cause fallacy, as there may be other factors that contributed to the student's poor performance, such as lack of preparation or understanding of the material.

\end{itemize}

Logical fallacies in the statement ``\textit{Caldwell Hall is in bad shape. Either we tear it down and put up a new building, or we continue to risk students’ safety. Obviously we shouldn’t risk anyone’s safety, so we must tear the building down}'':

\begin{itemize}
\item \textit{False Dilemma Fallacy}: The argument presents a false dilemma by suggesting that there are only two options: tearing down Caldwell Hall and putting up a new building or risking students' safety. This ignores the possibility of other solutions, such as renovating the existing building or relocating students to a different building.

\item \textit{Slippery Slope Fallacy}: The argument assumes that if we do not tear down Caldwell Hall, then we are automatically risking students' safety. This is a slippery slope fallacy, as there may be other ways to ensure student safety without tearing down the building, such as implementing safety measures or conducting regular inspections.

\item \textit{Hasty Generalization Fallacy}: The argument assumes that the state of Caldwell Hall is representative of all buildings on campus, or that all old buildings are in bad shape and pose a risk to students. This is a hasty generalization fallacy, as the state of one building does not necessarily represent the state of all buildings on campus.

\end{itemize}

\\

\cline{1-3}
\end{tabularx}
\label{table:topic}
\captionsetup{position=below}
\end{table*}

\begin{table*}[]
\small

\centering

\renewcommand{\arraystretch}{3}
\begin{tabularx}{\linewidth}{|T|Y|Z|}
\cline{1-3}

Supporting evidence generation & \begin{itemize} [leftmargin=*, itemindent=0pt, nosep]
    \item  \textbf{system role}: A helpful writing assistant focusing on argumentative essay tutoring and trying to suggest supporting evidence types in the given statement
    \item \textbf{user input}: Please list potential supporting evidences that can back the argument: [selected argument]. You can think from the following aspects: sharing professional experience (ethos), arousing audience’s emotion (pathos), providing facts and strict logical reasoning (logos) and presenting concrete practical
examples (example)
\end{itemize} & 

Supporting evidence types in the statement ``\textit{Renewable energy has the potential to significantly benefit people's lives in many ways. By reducing reliance on fossil fuels and transitioning to cleaner sources of energy, we can improve air quality and reduce the negative health effects associated with pollution. Additionally, renewable energy can create new job opportunities and boost local economies, particularly in rural areas where wind and solar energy projects can be developed}'':

\begin{itemize}
\item \textit{Statistical data (logos)}: Data from credible sources such as the International Energy Agency and National Renewable Energy Laboratory can provide statistical evidence of the potential benefits of renewable energy, such as reductions in air pollution and increases in job creation and economic growth.

\item \textit{Expert opinion (ethos)}: Opinions of experts, such as researchers and environmental scientists, can provide credibility to the argument that renewable energy can have significant benefits for people's lives.

\item \textit{Case studies (example)}: Examples of successful renewable energy projects, particularly in rural areas, can provide concrete evidence of the potential benefits of renewable energy.

\item \textit{Surveys and polls (logos)}: Surveys and polls can provide evidence of public opinion and support for renewable energy, as well as demonstrate the potential for consumer demand for renewable energy products and services.

\end{itemize}

\\

\cline{1-3}

Draft of key aspect & \begin{itemize} [leftmargin=*, itemindent=0pt, nosep]
    \vspace{\baselineskip}
    \item  \textbf{system role}: A helpful writing assistant focusing on argumentative essay tutoring. You are trying to write a starting sentence of the paragraph that support user's argument from a particular perspective
    \item \textbf{user input}: Write a starting sentence for the paragraph that elaborates on the argument [selected argument] from the perspective of [key aspect]
\end{itemize} & 
 N/A
\\
\cline{1-3}

Draft of discussion point & \begin{itemize} [leftmargin=*, itemindent=0pt, nosep]
    \vspace{\baselineskip}
    \item  \textbf{system role}: A helpful writing assistant focusing on argumentative essay tutoring. You are trying to elaborate on a particular given discussion point to support my argument.
    \item \textbf{user input}: Please write a paragraph that elaborates on my argument [selected argument] by considering the following discussion point [selected discussion point]
\end{itemize} & 
 N/A
\\
\cline{1-3}

\end{tabularx}
\label{table:topic}
\captionsetup{position=below}
\end{table*}

\begin{table*}[]
\small
\centering

\renewcommand{\arraystretch}{3}
\begin{tabularx}{\linewidth}{|T|Y|Z|}
\cline{1-3}

Draft of counter argument & \begin{itemize} [leftmargin=*, itemindent=0pt, nosep]
    \vspace{\baselineskip}
    \item  \textbf{system role}: A helpful writing assistant focusing on argumentative essay tutoring. You are trying to argue against an argument by considering a provided counter argument
    \item \textbf{user input}: Please write a paragraph that argues against the argument [selected argument] by considering the following counter argument [counter argument] from the perspective of [key aspect]
\end{itemize} & 
 N/A
\\

\cline{1-3}

Draft of supporting evidence & \begin{itemize} [leftmargin=*, itemindent=0pt, nosep]
    \vspace{\baselineskip}
    \item  \textbf{system role}: A helpful writing assistant focusing on argumentative essay tutoring. You are trying to support an argument by considering a provided supporting argument
    \item \textbf{user input}: Please write a paragraph that supports the argument: [selected argument] by realizing the following kind of supporting evidence [supporting evidence type]
\end{itemize} & 
 N/A
\\
\cline{1-3}

Settlement of logical fallacy & \begin{itemize} [leftmargin=*, itemindent=0pt, nosep]
    \vspace{\baselineskip}
    \item  \textbf{system role}: A helpful writing assistant. You are trying to fix the mentioned logical weaknesses in my argument
    \item \textbf{user input}: I just made an argument: [selected argument]. I know this argument has the following logical weaknesses: [logical fallacies]. Rewrite the argument to fix the logical weaknesses
\end{itemize} & 
 N/A
\\
\cline{1-3}

\end{tabularx}
\captionsetup{position=below}
\caption{Prompt templates used in goal recommendation, argumentative sparks and draft generation. According to the ChatGPT API specification, the "system role" determines the role that the model needs to assume during the current session, while the "user input" provides the model with the prompt that the user has inputted.}
\label{table:prompt_templates}
\end{table*}

\begin{table*}[]
\centering
\renewcommand{\arraystretch}{2.6}
\begin{tabularx}{\linewidth}{|m{\linewidth}|}
\cline{0-0}
\textbf{Topic2 [Attitudes and environment]:}
\newline
\textbf{\textcolor{figma_green}{People's attitudes are determined more by their immediate situation or surroundings than by society as a whole.}}
\newline
\textbf{\textcolor{figma_blue}{Write a response in which you discuss the extent to which you agree or disagree with the statement and explain your reasoning for the position you take. In developing and supporting your position, you should consider ways in which the statement might or might not hold true and explain how these considerations shape your position.}}\\

I disagree with the statement that people attitudes are determined more by their immediate situation or surroundings than by society as a whole. The two of these concepts go together, as what we are taught by society influences our reaction to immediate situations. For example, what you are taught from your parents and teachers when you are a child could determine how you handle situations you are in later in life. For example, I go to the store because I am hungry and need food. If I were to react based off of my attitudes of my immediate situation, I would be inclined to simply take the bag of chips and start eating because I am starving and want food. I have been taught by society that this is not right. The proper and appropriate thing to do would be to pick out the chips, pay for them, and then consume them. Therefore, society has taught me what is socially and morally acceptable and influences my decisions and attitudes more than my immediate situation. This can be applied to beliefs as well. Oftentimes, stereotypes that we are taught by society as children plays into our attitudes and beliefs as adults. This can have negative impacts, as seen in today's social climate. Regardless, what we are taught by society shapes our beliefs and attitudes more than our immediate situation. One could argue against this saying that as humans we have the capability to exercise free will and make our own decisions based off of the situation presented to us. In this view, what society has taught us does not matter because we can do whatever it is we want. For example, some people would go to the store and steal a bag of chips. Whether because they want to, or can't afford it. They know it is wrong. It is widely accepted by society that stealing is wrong. And yet some people exercise their free will to act off of their atitudes of the immediate situation: they are hungry, they steal the chips, their hunger is immedialy satisfied. In their view, the benefit of stealing the chips was greater than the cost, therefore prompting their attitudes to be shaped around their immediate situation rather than the views of society. But they had the develop this view from somewhere, probably a lesson from society (although a more negative lesson). Therefor, regardless of a positive or negative attitude, our attitudes are developed based off of society as opposed to our immediate situation because human thought is developed from somewhere, most likely society. \\
\cline{0-0}

\end{tabularx}
\label{table:topic_plain}\\
\captionsetup{position=below}
\caption{An example draft prototype generated by \textit{plain text editor} (i.e., baseline) from a user study session}
\end{table*}

\begin{table*}[]
\centering
\renewcommand{\arraystretch}{2.6}
\begin{tabularx}{\linewidth}{|m{\linewidth}|}
\cline{0-0}
\textbf{Topic1 [Scientific research funding]:}
\newline
\textbf{\textcolor{figma_green}{Governments should not fund any scientific research whose consequences are unclear.}}
\newline
\textbf{\textcolor{figma_blue}{Write a response in which you discuss your views on the policy and explain your reasoning for the position you take. In developing and supporting your position, you should consider the possible consequences of implementing the policy and explain how these consequences shape your position.
}}\\

 Governments should not fund any scientific research whose consequences are unclear.From an ethical standpoint, it is imperative that governments exercise caution when funding scientific research with uncertain consequences, as the potential risks to human life and the environment cannot be overlooked.Research findings are an essential part of scientific research, and they play a crucial role in shaping public policy and decision-making. However, it is important to recognize that research findings are not always clear-cut, and their interpretation can be subject to bias or misinterpretation. This is particularly true when dealing with scientific research that has uncertain consequences. In such cases, it is essential that governments exercise caution when interpreting research findings and consider the potential risks to human life and the environment. For instance, if a study finds that a particular chemical compound has potential therapeutic benefits, but also carries a significant risk of toxicity, the government must weigh the potential benefits against the risks before deciding to fund further research. Ultimately, it is the responsibility of the government to ensure that scientific research is conducted ethically and in a way that minimizes potential harm to both humans and the environment.  The discussion point "New Node" is an important consideration when it comes to funding scientific research with uncertain consequences. The development of new technologies and scientific breakthroughs can bring about significant advancements for society, but they can also pose potential risks to human life and the environment. Therefore, it is crucial for governments to exercise caution when providing funding for such research. The potential risks must be carefully evaluated and weighed against the potential benefits. Additionally, ethical considerations must also be taken into account, as the consequences of scientific research can have far-reaching implications for future generations. Ultimately, it is the responsibility of governments to ensure that the funding of scientific research is conducted in a responsible and ethical manner, with a focus on minimizing potential risks and maximizing benefits.  When it comes to scientific research, it is crucial to consider ethical principles before funding any project. Governments must exercise caution to ensure that the research aligns with ethical standards, especially when the consequences of the research are uncertain. Ethical principles require that the potential risks to human life and the environment are taken into account before any research is conducted. It is essential that the research does not harm human life, violate human rights, or negatively impact the environment. Governments must, therefore, ensure that the research is conducted with the utmost care and responsibility to prevent any adverse effects. In conclusion, ethical principles should be a top priority when funding scientific research, and the potential risks must be considered to ensure that the research does not have any negative consequences.  Case studies provide compelling evidence for the argument that governments must exercise caution when funding scientific research with uncertain consequences. For instance, the case of thalidomide, a drug developed in the 1950s to alleviate morning sickness in pregnant women, highlights the disastrous consequences of a lack of caution in scientific research. Despite insufficient testing, the drug was widely distributed, resulting in thousands of birth defects and deaths. Similarly, the case of genetically modified organisms (GMOs) illustrates the potential risks to the environment and human health that can result from insufficient testing and regulation. These examples demonstrate the importance of government oversight and caution in funding scientific research, particularly in cases where the consequences are uncertain or potentially harmful.  

Looking back at precedents of similar decisions in the past, it becomes clear that governments should not fund any scientific research whose consequences are unclear.  When it comes to funding scientific research, ethical considerations should always be at the forefront. While looking back at precedents of similar decisions in the past can be helpful in making informed decisions, it is important to remember that science and technology are constantly evolving. This means that the consequences of research may not always be entirely clear, even if we have historical examples to draw from. Therefore, it is crucial that governments carefully consider the ethical implications of any research they choose to fund. This includes considering the potential risks and benefits of the research, as well as the potential impact on vulnerable populations or the environment. Ultimately, funding scientific research should be done with the goal of advancing knowledge and improving society, but never at the cost of ethical principles.  The potential for unintended consequences of scientific research that is not fully understood is a crucial factor to consider when deciding whether or not to fund a particular research project. Looking back at precedents of similar decisions in the past, it is evident that many scientific discoveries have led to unforeseen negative consequences. For instance, the development of nuclear technology led to the creation of atomic bombs, which have caused widespread destruction and loss of life. Similarly, the widespread use of DDT as a pesticide led to environmental damage and health concerns. Therefore, it is essential to consider the potential risks and benefits of any scientific research before investing in it. Governments should prioritize funding research that has clear and well-understood consequences, rather than taking a gamble on research that could have unintended and harmful outcomes.  When it comes to funding scientific research, it is crucial to consider the potential consequences. If the outcomes of the research are unclear, it is difficult to determine the potential benefits or harms that may result. In these cases, the negative consequences of funding research with unclear consequences can be significant. For example, funding research that could lead to the development of dangerous weapons or technologies could have devastating consequences. Additionally, funding research that could harm the environment or public health without fully understanding the risks could also have severe negative consequences. Therefore, it is important for governments to carefully evaluate the potential consequences of funding scientific research before committing resources to it. By looking back at precedents of similar decisions in the past, we can see the importance of making informed decisions based on the potential consequences of the research.\textcolor{figma_orange}{\textbf{[continue in next page>>]}}  \\
\cline{0-0}

\end{tabularx}
\label{table:topic_plain}\\
\captionsetup{position=below}
\end{table*}

\begin{table*}[]
\centering
\renewcommand{\arraystretch}{2.6}
\begin{tabularx}{\linewidth}{|m{\linewidth}|}
\cline{0-0}
\textcolor{figma_orange}{\textbf{[>>continue for last page]}} While it is important to consider precedents of similar decisions in the past, it is also important to consider the potential for innovation. The counter argument suggests that funding scientific research with unclear consequences can lead to innovation. Many scientific breakthroughs and advancements have come from research that was initially deemed uncertain or risky. If governments only funded research with clear and immediate results, they would miss out on potentially groundbreaking discoveries. Additionally, it is difficult to predict the long-term consequences of any scientific research, so refusing to fund research with unclear consequences could limit progress and innovation in the future. Therefore, it is important to weigh the potential risks and benefits of funding scientific research and not solely rely on precedent as a deciding factor.  

From the perspective of potential risks of scientific research, it is imperative that governments exercise caution when funding research whose consequences are unclear.  When it comes to scientific research, it is essential to consider the potential risks associated with it, especially the potential harm to human health or the environment. Governments must exercise caution when funding research whose consequences are unclear, as the consequences can be catastrophic. The potential risks associated with scientific research can be numerous, including the release of hazardous chemicals or the creation of new diseases. In such cases, the consequences can be devastating, and the government must take measures to ensure that the risks are minimized. Therefore, it is crucial to evaluate the benefits and potential risks of the research before funding it, as it can have far-reaching consequences for the health and well-being of the population and the environment. Ultimately, it is the responsibility of the government to ensure that the research is conducted safely and that the risks are minimized to the greatest extent possible.  It is undeniable that scientific research has the potential to be misused by individuals or groups with malicious intent. This misuse could take many forms, from the development of new weapons to the creation of dangerous new viruses. Given this reality, it is essential that governments exercise caution when funding research whose consequences are unclear. In particular, they must carefully consider the potential risks associated with any research they choose to support, and take steps to mitigate those risks as much as possible. This may involve setting strict ethical and safety standards for researchers, or imposing regulations on the use of certain technologies. Ultimately, the goal should be to promote scientific progress while minimizing the potential for harm, and this requires a careful balancing of risks and benefits at all stages of the research process.  The difficulty of predicting the long-term consequences of scientific research is a major concern when it comes to funding scientific research. The potential risks of scientific research cannot be fully understood until the research has been conducted and the consequences become apparent. This is especially true for research that involves new technologies or untested hypotheses. Governments must exercise caution when funding research whose consequences are unclear, as the long-term effects of such research may be unpredictable and potentially harmful. Therefore, it is crucial that governments take a proactive approach in assessing the potential risks of scientific research before funding it, and ensure that researchers are conducting their studies in a responsible and ethical manner. By doing so, governments can help to minimize potential risks and ensure that the benefits of scientific research outweigh the potential harms. \\
\cline{0-0}

\end{tabularx}
\label{table:topic_GPT}\\
\captionsetup{position=below}
\caption{An example draft prototype generated by \textit{GPT Playground} from a user study session}
\end{table*}

\begin{table*}[]
\centering
\renewcommand{\arraystretch}{2.6}
\begin{tabularx}{\linewidth}{|m{\linewidth}|}
\cline{0-0}
\textbf{Topic3 [Career-oriented education]:}
\newline
\textbf{\textcolor{figma_green}{Educational institutions should actively encourage their students to choose fields of study that will prepare them for lucrative careers.}}
\newline
\textbf{\textcolor{figma_blue}{Write a response in which you discuss the extent to which you agree or disagree with the claim. In developing and supporting your position, be sure to address the most compelling reasons and/or examples that could be used to challenge your position.
}}\\

Educational institutions should actively encourage their students to choose fields of study that will prepare them for lucrative careers.Career development is a crucial aspect of a student's education, and educational institutions should actively encourage their students to choose fields of study that will prepare them for lucrative careers. By doing so, these institutions can help students achieve their financial goals and secure a stable future. Encouraging students to choose fields that offer high-paying jobs can also help address the issue of student debt. Students who graduate with high-paying jobs are better equipped to pay off their loans and avoid the financial burden that comes with student debt. Additionally, by promoting lucrative careers, educational institutions can contribute to the overall economic growth of a country. When students are trained in high-demand fields, they are more likely to find employment and contribute to the growth of various industries. Therefore, it is imperative that educational institutions prioritize career development and encourage students to consider fields that offer financial stability and growth opportunities. 
\newline 
One of the primary advantages of encouraging students to pursue fields of study that prepare them for lucrative careers is the economic benefits that students can enjoy throughout their lives.  Encouraging students to pursue fields of study that prepare them for lucrative careers not only benefits the students themselves but also their families. With a higher income, students can provide better financial support for their families, which can lead to an improved quality of life. For example, they can afford better housing, healthcare, and education for their children. Additionally, students can also contribute to their communities by investing in local businesses and philanthropic endeavors. By pursuing a career that offers a higher salary, students can not only secure their own financial stability but also positively impact the lives of those around them. Therefore, it is essential to encourage students to consider the economic benefits of their chosen field of study, not only for their own future but also for their families and communities.  When students pursue fields of study that prepare them for high-paying careers, they are more likely to be able to pay off their student loans and other debts in a timely manner. This is because they have the financial resources to do so, which can relieve a significant amount of stress and anxiety. The ability to pay off student loans and debt also allows students to invest in their futures by saving for retirement, purchasing a home, or starting a business. Furthermore, having a high-paying career can provide financial security and stability, which can lead to a better quality of life overall. Therefore, encouraging students to pursue fields of study that prepare them for lucrative careers can provide economic benefits that can last a lifetime.  Higher earning potential and career growth opportunities are two significant benefits of pursuing fields of study that prepare students for lucrative careers. By investing in education and training in high-demand fields such as computer science, engineering, finance, or healthcare, students can increase their earning potential significantly. These fields offer not only high starting salaries but also numerous opportunities for career advancement and growth. Students who pursue these careers can expect to earn more throughout their lives, which can lead to a more comfortable lifestyle and financial stability. Furthermore, the skills and knowledge gained in these careers can open doors to leadership positions and entrepreneurship, which can lead to even greater financial rewards. In short, encouraging students to pursue fields of study that prepare them for lucrative careers can provide them with lifelong economic benefits and opportunities for professional growth. \textcolor{figma_orange}{\textbf{[continue in next page>>]}}  \\
\cline{0-0}

\end{tabularx}
\label{table:topic_visar}\\
\captionsetup{position=below}
\end{table*}

\begin{table*}[]
\centering

\renewcommand{\arraystretch}{2.6}
\begin{tabularx}{\linewidth}{|m{\linewidth}|}
\cline{0-0}
\textcolor{figma_orange}{\textbf{[>>continue for last page]}} When it comes to choosing a field of study, practicality should be a key consideration for students looking to maximize their educational investment.  Career counseling services provided by educational institutions play a crucial role in helping students make practical and informed decisions when it comes to choosing a field of study. These services provide students with valuable information about the job market, industry trends, and the skills and qualifications required to succeed in various fields. By working with career counselors, students can assess their strengths and weaknesses, identify potential career paths that align with their interests and goals, and develop a plan to achieve their desired career outcomes. Such guidance helps students make informed decisions about their educational investments, ensuring that they choose a field of study that not only aligns with their interests but also offers them the best opportunities for career growth and financial stability. Therefore, students should take advantage of career counseling services provided by educational institutions to make practical and informed decisions when choosing their field of study.  The availability of internships and job placement programs should be a crucial factor when students are selecting their field of study. In today's competitive job market, acquiring relevant work experience is just as important as earning a degree. By choosing a practical field of study that offers ample internship opportunities and job placement programs, students can gain hands-on experience and develop a network of professional contacts that can help them secure employment after graduation. Additionally, internships can provide students with a glimpse into the day-to-day operations of their chosen industry, allowing them to make more informed decisions about their career path. Therefore, choosing a practical field of study that offers real-world experience can be a smart investment for students seeking to maximize their educational investment.  I completely agree with your argument that practicality should be a key consideration for students when choosing a field of study. However, it is not solely the responsibility of the students to ensure that their education aligns with the demands of the job market. Educational institutions also have a responsibility to equip their students with the skills and knowledge necessary to succeed in their chosen careers. This means that universities and colleges should offer programs that are tailored to the needs of the job market and provide opportunities for students to gain practical experience through internships, co-op programs, and other work-integrated learning opportunities. By doing so, educational institutions can help students maximize their educational investment and increase their chances of finding employment after graduation. Ultimately, both students and educational institutions must work together to ensure that graduates are well-prepared for the workforce and can make a positive impact on their chosen industries. 
\newline 
While it is true that the job market can be unpredictable, this is not a valid reason to shift the responsibility of practicality solely onto educational institutions. Students must also take responsibility for their own education and career prospects by researching job trends, networking with professionals in their desired field, and gaining relevant experience through internships or part-time jobs. Additionally, while the job market may be unpredictable, certain industries and fields consistently have high demand and job opportunities. Educational institutions should focus on providing programs and opportunities that align with these in-demand industries and equip students with the necessary skills to succeed in them. By solely relying on educational institutions to tailor their programs to the job market, students may miss out on valuable opportunities and fail to take proactive steps towards their own career success. Therefore, it is important for both students and educational institutions to work together and take responsibility for ensuring practicality in education. \newline So overall, our conclusion is educational institutions should actively encourage their students to choose fields of study that will prepare them for lucrative careers. \\
\cline{0-0}
\end{tabularx}
\label{table:draft_example}
\captionsetup{position=below}
\caption{An example draft prototype generated by \visar from a user study session}
\end{table*}

\begin{figure*}[ph!]
  \centering
  \includegraphics[width=\linewidth]{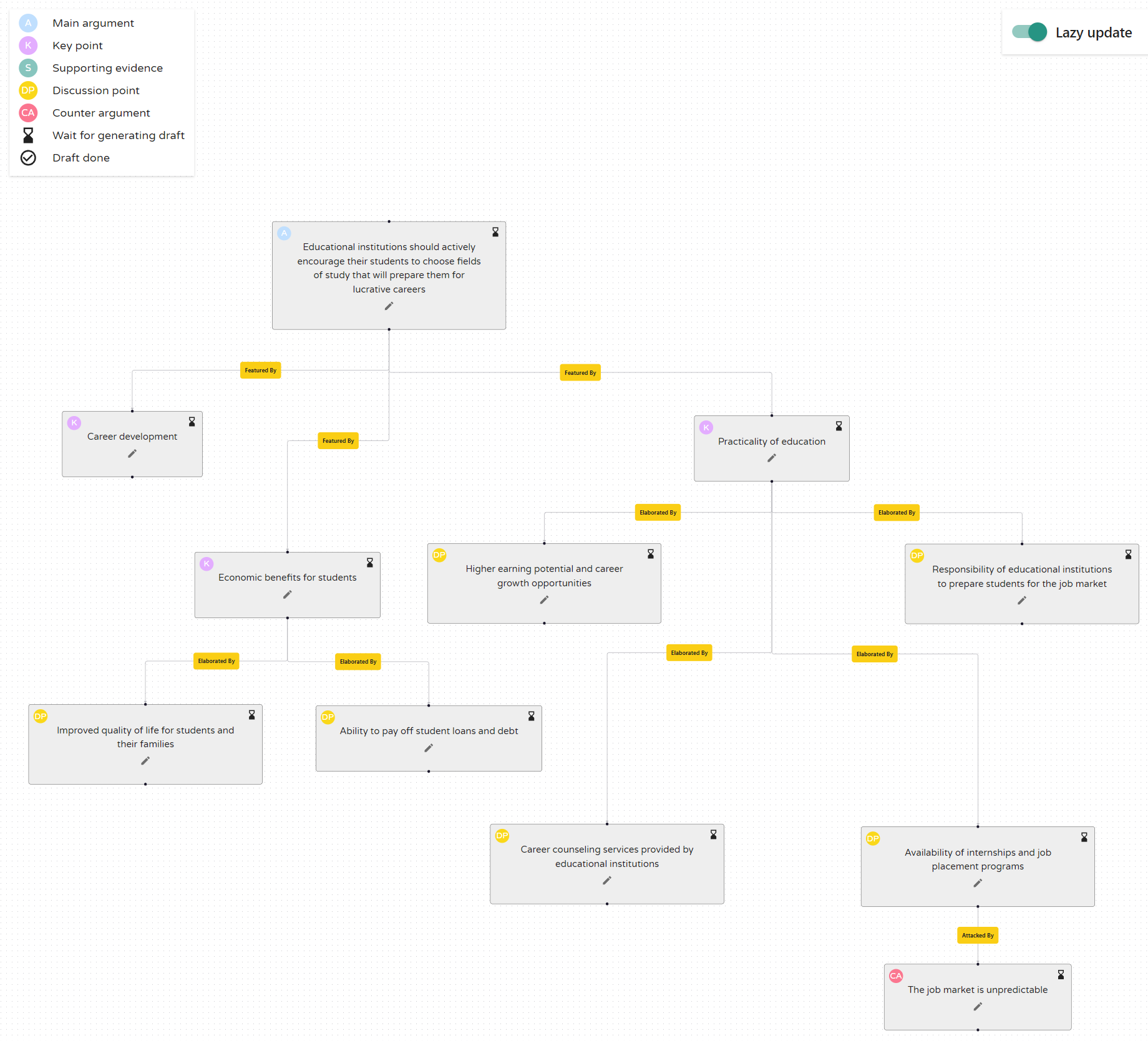}
  \caption{The visual outline of the example draft prototype shown in Table 3}
 
  \label{fig:example visual outline}
  \vspace{-3mm}
\end{figure*}

\end{document}